\documentclass[reprint,amsmath,amssymb,aps,prb,floatfix,a4paper,superscriptaddress]{revtex4-1}

\usepackage{graphicx}% Include figure files
\usepackage{dcolumn}% Align table columns on decimal point
\usepackage{bm}% bold math
\usepackage{color}
\usepackage[ugly]{nicefrac}
\usepackage[latin1]{inputenc}
\usepackage{subfig}
\usepackage{SIunits}
\usepackage{epstopdf}
\newcommand{\mrm}[1]{$\mathrm{{#1}}$}

\newcommand{\etal}{$\textit{et\ al.\ }$}

\begin{document}
\title{Structural, chemical and electronic properties of the $\mathrm{Co_{2}MnSi(001)/MgO}$ interface}
\author{Roman Fetzer}
\email{rfetzer@rhrk.uni-kl.de}
\author{Jan-Peter W\"ustenberg}
\author{Tomoyuki Taira}
\author{Tetsuya Uemura}
\author{Masafumi Yamamoto}
\affiliation{Division of Electronics for Informatics, Hokkaido University, Kita 14 Nishi 9, Sapporo 060-0814, Japan}
\author{Martin Aeschlimann}
\author{Mirko Cinchetti}
\affiliation{Department of Physics and Research Center OPTIMAS, University of Kaiserslautern, Erwin-Schr\"odingerstr.\,46, 67663 Kaiserslautern, Germany}
\date{\today}
%%%%%%%%%%%%%%%%%%%%%%%%%%%%%%%%%%%%%%%%%%%%%
\begin{abstract}

The performance of advanced magnetic tunnel junctions build of ferromagnetic (FM) electrodes and MgO as insulating barrier depends decisively on the properties of the FM/insulator interface. Here, we investigate interface formation between the  half-metallic compound $\mathrm{Co_{2}MnSi}$ (CMS) and MgO  by means of Auger electron spectroscopy, low energy electron diffraction and low energy photoemission. The studies are performed for different annealing temperatures ($T_{A}$) and MgO layer coverages (4,~6,~10,~20 and 50~ML). Thin MgO top layers ($t_{MgO}\leq$10~ML) show distinct surface crystalline distortions, which  can only be partly healed out by annealing and furthermore lead to distinct adsorption of carbon species after the MgO surface is exposed to air. For $t_{MgO}>$~10~ML the MgO layer surface exhibits clearly improved crystalline structure and hence only marginal amounts of adsorbates. We attribute these findings to MgO misfit dislocations occurring at the interface, inducing further defects throughout the MgO layer for up to at least 10~ML. Furthermore, spin-polarized photoemission spectra of the CMS/MgO interface are obtained for MgO coverages up to 20~ML, showing a clear positive spin polarization near the Fermi energy ($E_F$) in all cases. 
\end{abstract}
%%%%%%%%%%%%%%%%%%%%%%%%%%%%%%%%%%%%%%%%%%%%%
\maketitle
%%%%%%%%%%%%%%%%%%%%%%%%%%%%%%%%%%%%%%%%%%%%%
\section{Introduction}
\label{sec:intro}

%Spintronics, MTJs, CMS
Magnetic tunnel junctions (MTJs) are prototypical spintronics devices, consisting of two ferromagnetic leads separated by an insulating barrier~\cite{Moodera95, Tsymbal03, Zutic04, Velev08}. As first shown by Julliere, the resulting tunneling magnetoresistance (TMR) depends crucially on the electron spin polarization  of the electrodes~\cite{Julliere75}. Therefore effectiveness of such devices can be increased significantly by using highly spin polarized half-metallic  materials. Amongst other predicted half-metals, the Heusler compound $\mathrm{Co_2MnSi}$ (CMS) features a high Curie temperature of 985~K and proper epitaxial growth~\cite{Brown00, Kaemmerer04, Ishikawa06}. The predicted minority band gap~\cite{Ishida98, Balke06, Chadov09} was experimentally demonstrated by tunneling spectroscopy at low temperature, having a width of aprox. 400~meV, where the Fermi energy ($E_F$) lies very close to the conduction band minimum~\cite{Sakuraba06b, Ishikawa06}. Later studies confirmed the band gap width but found $E_F$ in the middle of the minority band gap~\cite{Ishikawa09a}. Indeed a superior performance of CMS compared to conventional 3d ferromagnetic materials in MTJs using an amorphous $AlO_x$ barrier could be demonstrated at least for low temperatures~\cite{Sakuraba06a}. Further improvement was realized by using epitaxially grown MgO instead of $AlO_x$ as the tunnel barrier~\cite{Ishikawa06, Ishikawa08}. As Butler \etal first predicted for Fe/MgO/Fe MTJs, the tunneling probability will depend additionally on the electron wavefunction symmetry in case of crystalline barriers due to $k_{||}$-conservation~\cite{Butler01}. Using ferromagnetic materials with appropriate band structure, the preferential tunneling leads to significantly increased TMR ratios. This also holds for CMS/MgO/CMS MTJs~\cite{Miura07}. Ishikawa and co-workers could raise the TMR value further on even above 1000\% by varying the chemical composition of CMS~\cite{Ishikawa09b}. \newline
Bulk MgO possesses a direct band gap at the $\Gamma$ point~\cite{Chang84, Xu91}. The band gap width is about 7.8~eV~\cite{Roessler67, Pandey91}, but can be reduced by defect induced gap states~\cite{Gibson94,Kantorovich95,Illas98} as well as surface states~\cite{Lee78, Schoenberger95}. For MgO thin films the band gap width generally is reduced and depends on fabrication procedure ~\cite{Mather06, Sterrer06b, Afanasev10, Klaua01, Dedkov06}. However, none of these cases leads to a finite density of states directly ($\pm$ 0.5~eV) at $E_F$, which usually falls in the middle of the MgO band gap.\newline
Epitaxial growth of MgO on top of CMS thin films inevitably leads to misfit dislocations at the interface due to the relatively large lattice mismatch of 5.1\%~\cite{Miyajima09}. Oxidation, as broadly discussed for the Fe/MgO interface, does not take place at the CMS/MgO interface in case of electron beam evaporation of the MgO layer~\cite{Saito07}.
In general, the performance of MTJs depends critically on heat treatment. Usually an optimum annealing temperature ($T_A$) for such devices has to be found, which results in a maximum TMR value ~\cite{Jourdan09}. For devices consisting of CMS and MgO, the optimum annealing temperature is in the range between 450 and 600°C~\cite{Ishikawa09b, Sakuraba06a, Sakuraba10}. \newline
A lot of effort has been put in the past into characterization of MgO thin films grown on different metals. For example, Auger electron spectroscopy (AES) was performed at MgO thin films grown on metals in order to find out the surface chemical composition, i.e. wether stoichiometry is given~\cite{Valeri01} or interdiffusion takes place at the interface~\cite{Vassent96}. \newline
Other detailed investigations were done by low energy electron diffraction (LEED) for Fe/MgO bilayers~\cite{Dynna96, Wulfhekel01, Klaua01}. Pseudomorphic growth of MgO for up to 5-7~ML was found, resulting in sharp LEED spots. By overcoming the critical thickness, lattice mismatch (3.5\%) induced misift dislocations occur at the interface, which lead to warped and tilted surface segments and increased surface roughness. In this case the obtained LEED patterns consisted additionally of satellite spots with fourfold symmetry. For thicker MgO top layers the tilting at the surface decreases, therefore satellite and main spots smear out to single broad spots. \newline 
Regarding the characterization of the spin-dependent electronic properties, up to now only the bare surface of different  Heusler compounds has been investigated by spin-resolved photoemission spectroscopy (SR-PES)~\cite{Wang05a,Wuestenberg12, Cinchetti07, Schneider06, Wuestenberg09, Kolbe12}. Applying this experimental technique to MTJ interfaces was done so far only for Fe/MgO, CoFe/MgO and Co/MgO bilayers in order to reveal the electronic properties regarding spin and symmetry~\cite{Sicot03, Matthes04, Mueller07, Tong06, Bonell12}. Due to the large band gap of MgO, Fe bulk states and possible interface states are in principle observable near $E_F$. However, in these studies the MgO coverage could not exceed more than 2~ML. For thicker MgO coverage, due to the high surface sensitivity  of conventional photoemission, the photoemission yield stems mainly from MgO. This furthermore results in a distinct drop of the detected surface spin polarization (SP)~\cite{Dedkov06, Plucinski07}. Hence the results can hardly be applied to MTJ properties, where the tunneling barrier usually has a thickness of at least 6~ML. Also pinholes can not be excluded for such thin MgO layers ($\leq 2$~ML), leading to a certain amount of metal surface contribution to the photoemission signal. \newline
Here we investigate comprehensively the formation of the interface of advanced TMR devices consisting of halfmetallic CMS and insulating MgO by means of LEED, AES and low energy SR-PES. These methods are applied to a set of CMS/MgO samples with varying MgO top layer thickness ($t_{MgO}$) ranging from 4~ML to 50~ML. For thin MgO coverages ($t_{MgO}\leq10$~ML) we find significant deviations of the crystalline ordering and elemental composition of the MgO surface. The spin-dependent electronic properties of the CMS/MgO interface are investigated directly via spin-resolved photoemission. This is possible by using a very low photon energy of $h\nu=5.9~eV$~\cite{Fetzer12b}. A distinct positive spin polarization at $E_F$ is found for MgO top layer thicknesses up to 20~ML, which drops monotonically for higher binding energies. All measurements on each of the samples were conducted in dependence of the annealing temperature, ranging from 400 to 600°C. We find that, although the chemical composition of the MgO layers does not vary by annealing, strong changes appear for the thinner samples in crystal structure and photoemission spectra, again pointing out the influence of interface defects to the MgO top layer.

%%%%%%%%%%%%%%%%%%%%%%%%%%%%%%%%%%%%%%%%%%
\section{Samples}
\label{sec:samples}

The investigated samples were grown epitaxially in an ultrahigh vacuum chamber with a base pressure of \unit{6\times 10^{-10}}{mbar}~\cite{Yamamoto10}. The stack structure was the following: MgO(001)sub/MgO(10~nm)/CMS(30~nm)/MgO(x~nm); x~=~4,~6,~10,~20 and 50~ML with 1~ML~$\cong$~0.211~nm. As the minimum MgO thickness we have chosen $t_{MgO}=4$~ML ($0.8$\,nm) since for thinner layers pinholes might occur. This would affect the investigated sample properties drastically, i.e. oxidation of the Heusler layer would take place when transferring ex-situ. The CMS layer was evaporated by magnetron sputtering at room temperature (RT) and afterwards annealed up to 600°C. The bulk chemical composition deviates from the stoichiometric case with an actual Mn-rich composition of  \mrm{Co_2Mn_{1.19}Si_{0.88}}. MTJs using such Mn-rich compositions show distinctively improved TMR values up to a certain amount of Manganese ~\cite{Ishikawa09b}. The MgO(001) top layers were deposited by electron beam evaporation at RT. No additional annealing took place in the evaporation chamber. The samples were transferred ex situ to the experimental chamber, immersed into a non-water-containing liquid since MgO is known to be hygroscopic. \newline  
Prior to the conducted experiments, the samples were annealed a second time via resistive heating in several steps from 400°C up to 600°C for at least 20 min, which allowed for systematic study of the dependence of the properties of the MgO layer as well as of the CMS/MgO interface on the annealing temperature ($T_A$). The actual temperature $T_A$ was controlled by a calibrated Pyrometer measuring the sample holder surface temperature, with the Pyrometer spot directly beside the sample. The optimum annealing temperature for the CMS/MgO system is known to be 600°C~\cite{Ishikawa09b, Yamamoto10}.

%%%%%%%%%%%%%%%%%%%%%%%%%%%%%%%%%%%%%%%%%%
\section{Experimental set-up}
\label{sec:exp}
%% LEED
The setup of the experimental chamber is identical to the one described in detail by Wüstenberg \etal~\cite{Wuestenberg12}. LEED patterns of the samples were obtained by using a 3-grid SpectaLEED system manufactured by Omicron. In all cases a primary electron energy~($E_P$) of 90~eV was used to ensure high surface sensitivity. This allowed us to investigate the crystalline ordering degree of the outermost MgO surface layers. \newline %%Auger
An Omicron CMA 100 energy analyzer in combination with an electron gun ($E_P = 3keV$) was used for Auger electron spectroscopy, revealing the relative chemical composition of the MgO top layer surface~\cite{Davis87}. For every species a characteristic peak at a specific kinetic energy is evaluated, i.e. peak height and sensitivity factor (depends on element and energy) are taken into account to determine the element-resolved surface composition. \newline %% SR-PES Setup
The excitation source used for the photoemission experiments was the 4th harmonic of a Spectra Physics Tsunami Ti:Sapphire oscillator  with  photon energy of 5.9~eV. The laser light was linearly p-polarized with an angle of incidence of 45° with respect to the surface normal. Photoelectrons leaving the surface in normal emission, i.e. the $\Gamma$-X direction, were detected by a 90° Focus CSA 300 cylindrical sector analyzer with an energy resolution of 210~meV~\cite{Wuestenberg12}. Since the perpendicular component of the electron wave vector is not conserved during photoemission, the whole $\Gamma X$ part of the Brillouin zone which is accessible with the used photon energy contributes to the photoemission spectrum. Only in this part a finite density of states at the Fermi energy is predicted~\cite{Miura07, Balke06}. Additionally, due to the finite acceptance angle of our analyzer and further usage of a biasing voltage, we probe approx. 60\% of the Brillouin zone in $\Gamma$-K direction~\cite{Wuestenberg12}. After energy detection an additionally mounted Focus SPLEED analyzer is used to determine the spin asymmetry of the incoming photoelectrons. The actual spin polarization is calculated by using a Sherman factor value of 0.2. All experiments were carried out at room temperature.

%%%%%%%%%%%%%%%%%%%%%%%%%%%%%%%%%%%%%%%%%%%%%%%%%%%%%%%%%%%%%%%%%%%%%

\section{AES results}
\label{sec:AES}

\begin{figure}
\centering
\includegraphics[width=\linewidth]{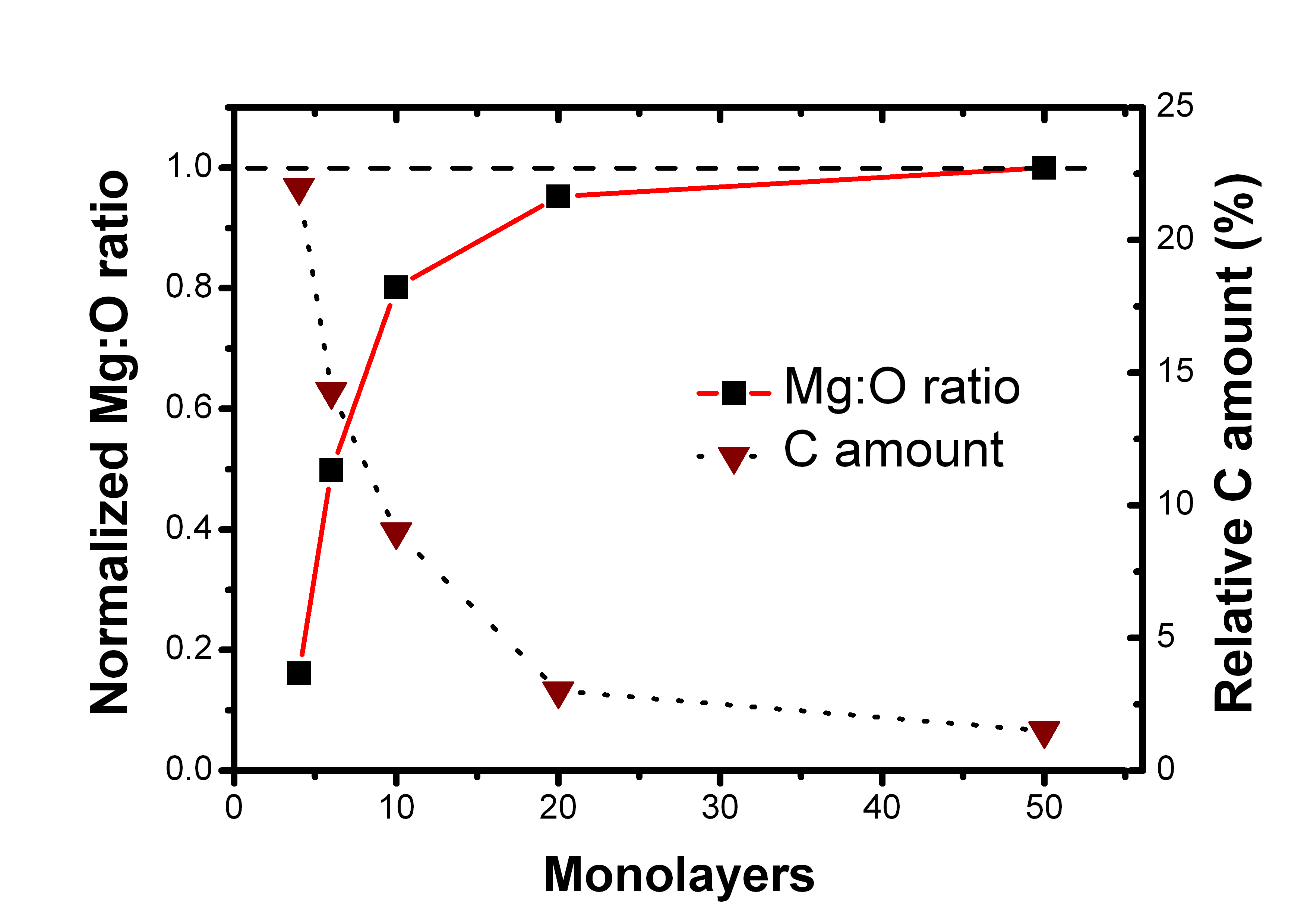}
\caption{Mg:O ratio and relative carbon amount for varying MgO top layer thickness for CMS/MgO samples annealed at $T_A=600°C$. The connecting lines serve as a guide to the eye.}
\label{fig:AES}
\end{figure}

The AES spectra (not shown here) are in general dominated by Mg and O signal. For thinner MgO top layers, the elements of the underlying Heusler compound Co, Mn and Si also contribute. For none of the samples a distinct change of the chemical composition in dependence of the annealing temperature is found, which means a significant interdiffusion at the interface does not take place at the investigated temperature range. Figure \ref{fig:AES} shows the Mg:O ratio in dependence of the MgO top layer thickness. The measured values are normalized to the bulk Mg:O ratio which was obtained at a well-sputtered and heated MgO substrate. The data in Fig. \ref{fig:AES} are obtained after annealing to 600°C.  While the samples with $t_{MgO}=$20~ML and 50~ML  show almost stoichiometric composition, the detected relative Mg amount decreases monotonically with respect to the oxygen signal for thinner coverages, reaching a minimum normalized ratio value of 16\% for the sample with $t_{MgO}=4$~ML. Valeri \etal observed a similar decrease of th Mg:O ratio when reducing the film thickness on top of silver from 20~ML down to 1~ML. They explained these findings by considering different electron inelastic mean free paths (IMFPs) for the evaluated O~KLL (503~eV, $\lambda_{IMFP}\approx13~\mathring{A}$) and Mg~KLL (1174~eV, $\lambda_{IMFP}\approx25~\mathring{A}$) peaks~\cite{Valeri01, Nist00}. Therefore the Mg signal apparently decreases because the IMFP is in the range of the film thickness. Taking this effect into account, the right stoichiometry could be shown also for the thinnest MgO layers. However, this is not the case in our experiments, since even for a MgO thickness of only 1~ML the  apparent reduction of the Mg:O ratio due to IMFP effects would not exceed 50\%, in contrast with our results. \newline
Additionally to the expected species, carbon was found on the surface of every sample, which probably originates from adsorption of carbon oxide molecules while the samples are exposed to air. This is inevitable since they are transferred ex situ to the experimental chamber. Remarkably, the relative C amount on top of the samples depends unambigiously on the MgO thickness, as can be seen from Fig. \ref{fig:AES}. For samples with $t_{MgO}$~=~20~ML and 50~ML the amount of C is neglectable (values lower than 3\%), while a clear monotonic increase occurs when going to lower MgO coverages resulting in a carbon amount of more than 20\% for the thinnest MgO top layer thickness of 4~ML. It is well known that the perfect MgO surface is almost inert to molecular oxygen and carbon oxides, while adsorption of these species is likely to occur if point defects, steps or terraces are present at the surface~\cite{Henrich85, Geneste05, Ochs98, Pacchioni94, Pacchioni93, Pacchioni92, Kantorovich96, Kantorovich97}. The enrichment of carbon oxides and molecular oxygen at the surface would furthermore result in an enhanced O signal, hence leading to the observed decrease of the Mg:O ratio. \newline
In conclusion, our observations point to a higher defect density at the MgO surface for lower MgO coverages. This behavior can be ascribed to misfit dislocations occuring at the CMS/MgO interface~\cite{Miyajima09}, which can propagate to or at least induce defects and disorder at the MgO surface of samples with low MgO layer thickness, while for higher coverages the influence vanishes.

%%%%%%%%%%%%%%%%%%%%%%%%%%%%%%%%%%%%%%%%%%%%%%%%%%%%%%%%%%%%%%%%%%%%%

\section{LEED results}
\label{sec:LEED}

\begin{figure}
\centering
\includegraphics[width=\linewidth]{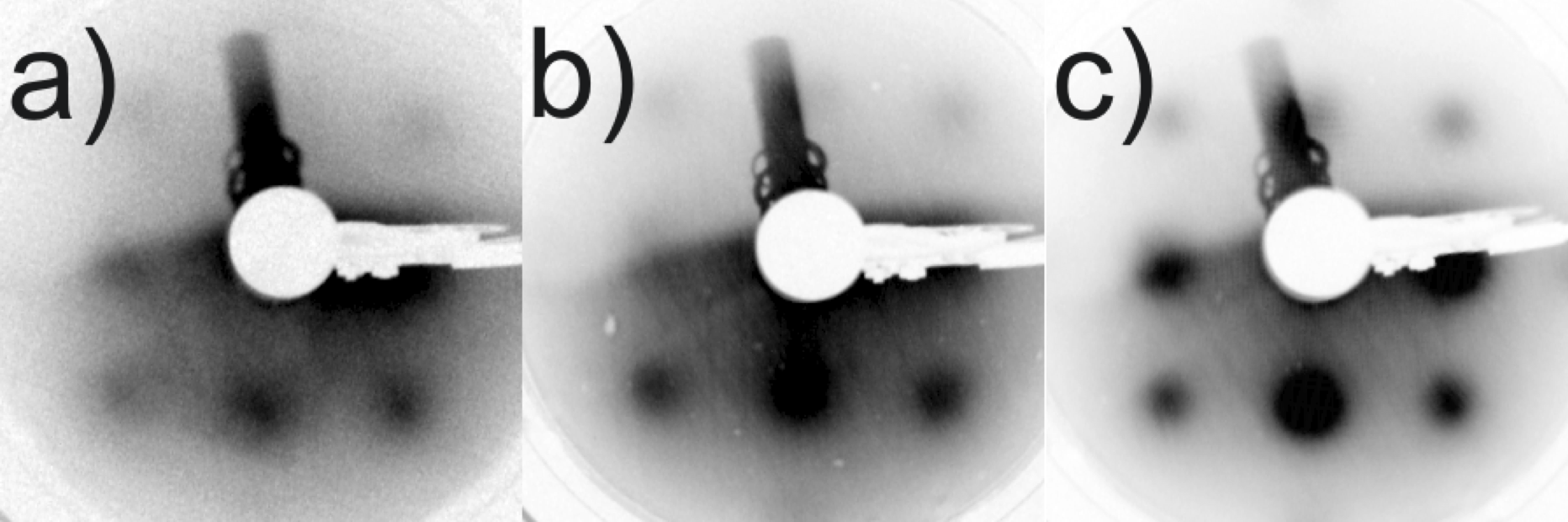}
\caption{LEED patterns at a primary electron energy of 90~eV of a) $t_{MgO}$~=~10~ML annealed to 400°C,  b) $t_{MgO}$~=~10~ML annealed to 600°C and c) $t_{MgO}$~=~20~ML annealed to 400°C. Please note that the upper half of the luminescent screen is less sensitive.}
\label{fig:LEED}
\end{figure}

%%%%%%%%%%%%%%%%%%%%%%%%%%%%%%%%%%%%%%%%%%%%%%%
LEED experiments were conducted on all samples in dependence on the (second) fannealing temperature $T_A$(400°C, 500°C, 600°C). 
For the sample with the thinnest MgO barrier (4~ML) a LEED  pattern could not be observed, independently of $T_A$. The reason for the absence of diffraction spots may be the large amount of adsorbed carbon on the MgO surface as found by AES (c.f. section \ref{sec:AES}). However, one has to keep in mind that the adsorption itself probably occurs due to a large defect density at the surface, which would also hinder the formation of a LEED pattern.
For the samples with $t_{MgO}>4$~ML we could observe the  LEED pattern expected from the MgO B1 crystal structure~\cite{Wuestenberg12}. 
In particular, for the samples with $t_{MgO}=$6~ML and 10~ML we observed a significant improvement of the pattern quality by increasing $T_A$. This is illustrated in Fig. \ref{fig:LEED} a) and b), showing LEED patterns obtained from the sample with $t_{MgO}=10$~ML annealed to $T_A$=400°C and $T_A$=600°C, respectively. Here, an improvement of the pattern quality for higher annealing temperature is clearly visible. Since the chemical composition does not vary with $T_A$ (c.f. section \ref{sec:AES}), we attribute this effect solely to the occurrence of defect healing throughout the whole MgO layer. Nevertheless, even for $T_A$=600°C the spots are far from being sharp probably due to the presence of residual defects and adsorbed molecules at the surface. 

\noindent In contrast to the previous behavior, the LEED patterns of the samples with $t_{MgO}=$20~ML and 50~ML are hardly influenced by annealing. The obtained patterns have better quality compared to those recorded for lower MgO coverages, showing smaller spot widths and recognizable intensity differences.  Figure \ref{fig:LEED} c) shows exemplarily the LEED pattern obtained from the sample with $t_{MgO}=$20~ML and $T_A=400$°C, demonstrating that the LEED pattern quality is enhanced already for low $T_A$. Here the spot sizes are comparable to those reported in several other publications on non-magnetic substrates and different MgO layer thicknesses~\cite{Wu91, Tegenkamp99, Valeri01}. Again this can be explained by a diminished influence of the interface misfit dislocations to the sample surface in case of higher MgO coverages, since less surface defects lead to an improved LEED pattern quality, while annealing effects are reduced. \newline 
However, neither point-like spots for very thin MgO layers induced by pseudomorphic growth nor additional satellite spots due to misfit dislocations occuring at thicker MgO coverages are found. This is in contrast to the protoype MTJ interface Fe/MgO~\cite{Dynna96, Klaua01, Wulfhekel01}, which should be well comparable do the CMS/MgO interface since the lattice mismatch is similar and has the same sign. Our measurements suggest that pseudomorphic growth at the interface can only occur for $t_{MgO}<$~4ML. In fact, for this type of growth the MgO surface should be almost perfectly ordered and therefore no adsorption should take place. The satellite spots would be only observable in a certain layer thickness region, after exceeding the critical thickness and before the satellite spots and main spots blur due to the lowered surface tilting. The latter case can not be excluded for the thicker MgO coverages of 20~ML and 50~ML (c.f. Ref~\cite{Dynna96}). For the thinner MgO thicknesses it is not possible to say if satellite spots occur or blurring takes already place since the LEED pattern quality is too low due to other surface defects and/or adsorbed species. 
Before passing, we would like to stress that our results, indicating an imperfect  MgO surface for typical MgO thickness used in TMR elements (6-10~ML) explain the fact that in TMR elements the upper CMS electrode usually grows worse compared to the lower one, leading to a decreased minority band gap width and to residual bulk states in the gap itself~\cite{Ishikawa09a}.

%%%%%%%%%%%%%%%%%%%%%%%%%%%%%%%%%%%%%%%%%%%%%%%%%%%%%%

\section{SR-PES results}
\label{sec:PES}

\begin{figure}
\centering
\includegraphics[width=\linewidth]{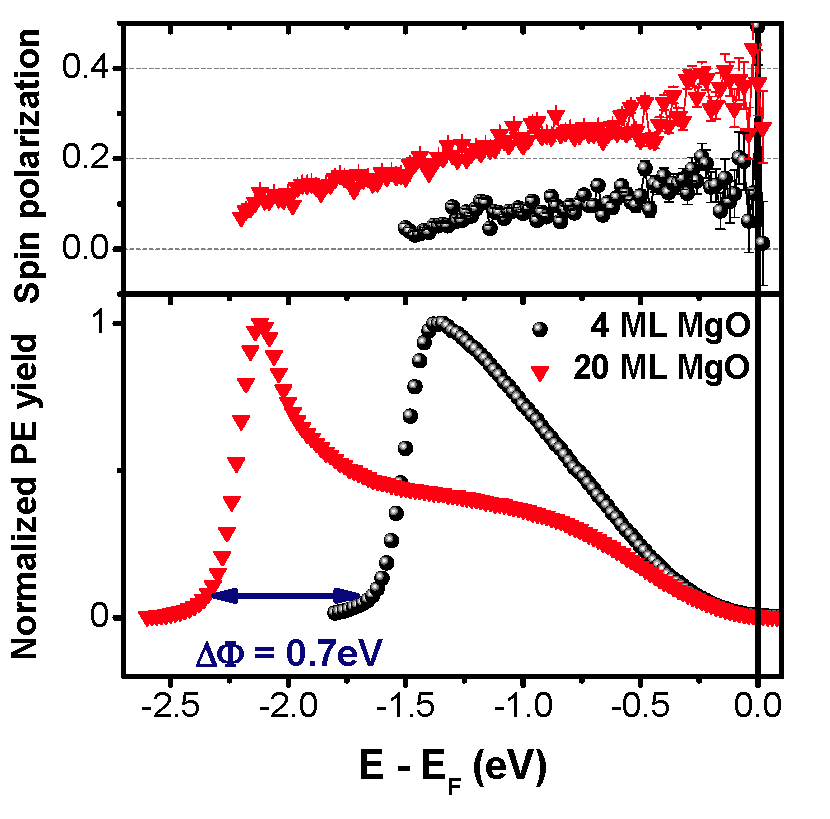}
\caption{Spin-integrated and normalized spectra (lower part) and spin polarization (upper part) for samples with $t_{MgO}$~=~4 and 20~ML with $T_A$~=~600°C.}
\label{fig:PES1}
\end{figure}

The work function of all samples is lower than the used photon energy of 5.9~eV, which in principle allows us to obtain spin-resolved photoelectron spectra in all cases. Although MgO is insulating, only the sample with the thickest MgO coverage of 50~ML shows charging effects, which can not be overcome by reducing the laser intensity. Therefore this sample will not be considered henceforth in this paragraph. We will start with a comparison of the results obtained after annealing to the optimal temperature of $T_A=$600°C. \newline
Figure~\ref{fig:PES1} shows a comparison of the normalized (spin integrated) spectra (bottom panel) and the corresponding  spin polarization (SP, upper panel) for the samples with $t_{MgO}=$4~ML and 20~ML ($T_A=$600°C). Both samples with intermediate MgO thickness ($t_{MgO}$~=~6~ML and 10~ML) and $T_A$~=~600°C revealed spectra and SP very similar to the one with $t_{MgO}$~=~4~ML. 
The data in Fig.~\ref{fig:PES1} show two very surprising facts: first of all, a non-vanishing spin polarization is detected  despite the presence of a non-magnetic MgO layer with thickness up to 20~ML on top of CMS; second, the photoemission yield does not drop significantly for thicker MgO coverages. The origin of this unexpected behavior is the drastically reduced phase space for inelastic scattering of electrons excited at the  CMS/MgO  interface with MgO valence electrons, because their excess energy ($\leq h\nu=$5.9~eV) is smaller than the MgO band gap energy of 7.8~eV~\cite{Roessler67}. As a result, the photoelectrons excited at the CMS/MgO interface do not lose significantly kinetic energy when traversing the MgO layer. This makes low energy SR-PES sensitive to interfaces buried below insulator layers, as recently demonstrated in  Ref.~\cite{Fetzer12b}.

\noindent Comparing the spin-integrated interface spectra in Fig.~\ref{fig:PES1} to those of the free off-stoichiometric CMS surface in Ref.~\onlinecite{Wuestenberg12} and those from the free stoichiometric CMS surface Ref.~\onlinecite{Fetzer12b}, we observe that in the CMS/MgO interface spectra the spectral features are washed out.
This is due to inelastic scattering at defects mostly at the very MgO surface region. This effect is even more pronounced than previously reported in Ref.~\onlinecite{Fetzer12b}, as in contrast to Ref.~\onlinecite{Fetzer12b} here the MgO surface was not sputter-cleaned before the photoemission experiments. Scattering at the MgO surface affects almost all excited electrons and leads to a diminished photoemission yield at the Fermi edge. At lower energies the spectra and the resulting spin polarizations are significantly influenced by secondary electrons which lost kinetic energy through scattering, hence leading to a monotonic increase of the photoemission yield and monotonic decrease of the SP. For this reason, only the spin polarization directly at $E_F$ resembles the true interface spin polarization. Here we always find a distinct positive value, in contradiction to the theoretical investigations of Miura \etal, who report large additional minority electron density at the CMS/MgO interface induced by interface states~\cite{Miura08}. This results in a very low or even negative interface SP (however, these states do not influence the TMR ratio, since they are not able to couple to bulk Bloch states). The discrepancy may be explained by the symmetry selection rules valid for photoemission, which usually do not allow excitation of all energetically available states~\cite{Hermanson77, Eberhardt80}. 
Strikingly the SP at $E_F$ for $t_{MgO}$~=~20~ML and $T_A$~=~600°C has a value of almost 40\%,  a factor of two higher than for $t_{MgO}$~=~4~ML. This has its origin in different properties of the outermost MgO layers, since the CMS and interface properties are the same for all samples. This is evidenced by a large work function difference $\Delta \Phi$ of 0.7~eV as depicted in the left part of Fig.\ref{fig:PES1}, since $\Phi$ is a unique property of the surface. Therefore we attribute the decreased SP in case of thinner MgO coverage to additional non-spin-conserving scattering at surface defects and adsorbed molecules which were discussed in the previous sections.
 \begin{figure}
\centering
\includegraphics[width=\linewidth]{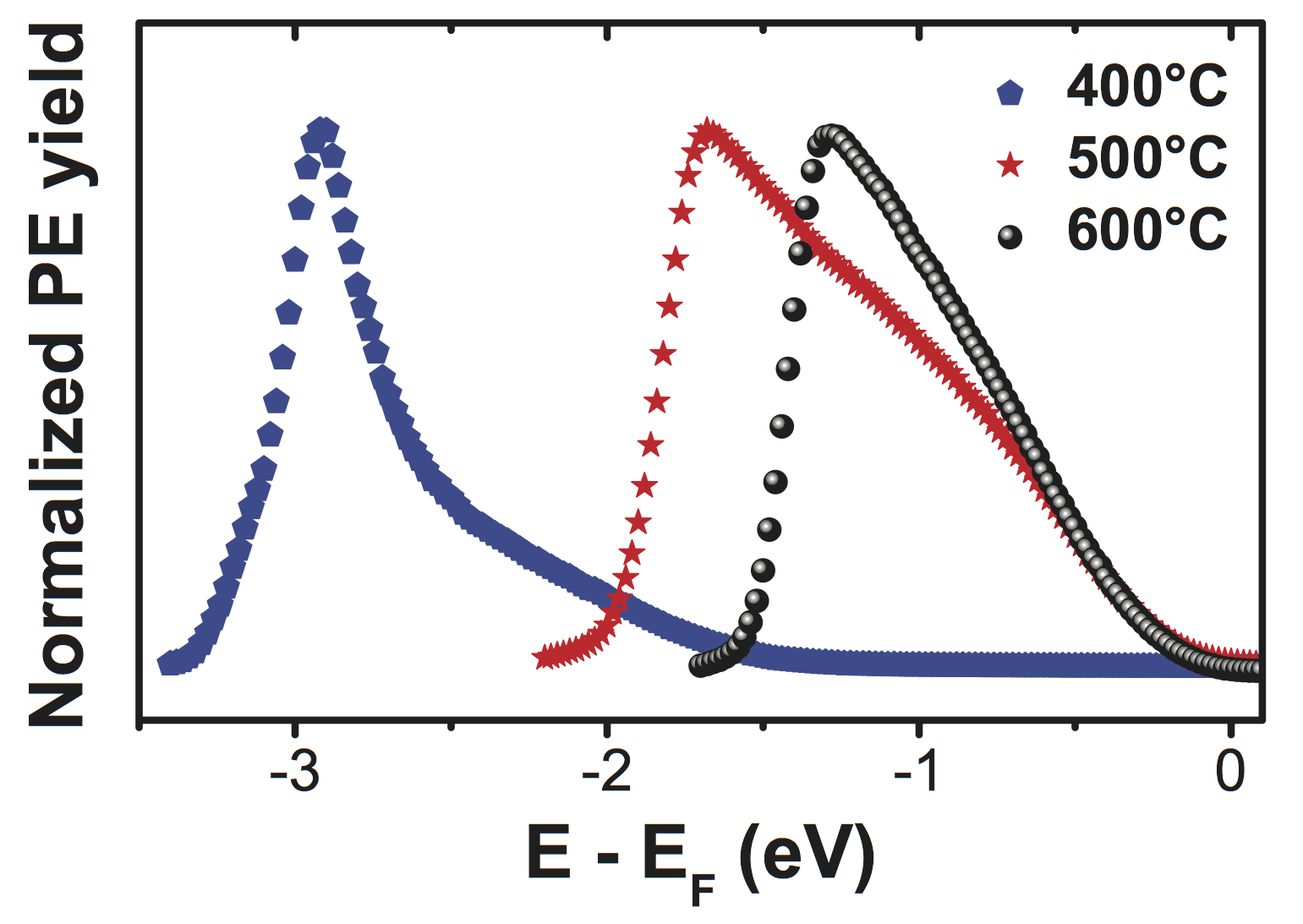}
\caption{Spin-integrated and normalized spectra for $t_{MgO}$~=~4~ML in dependence of $T_A$.}
\label{fig:PES2}
\end{figure}

\noindent Additionally $T_A$-dependent studies are performed in the interesting temperature regime of 400-600°C. Spectra and resulting SP (not shown) do not change in the case of $t_{MgO}$~=~20~ML, as it was the case for the observed LEED patterns in section \ref{sec:LEED}. Quite contrary, both samples with intermediate MgO coverage ($t_{MgO}$~=~6 and 10~ML, not shown) show a  monotonic but rather small workfunction increase by in total 0.35~eV and 0.15~eV, again in correspondence to our observations made by LEED. However the shapes of the photoemission spectra near $E_F$ do not alter, and correspondingly the SP stays constant within the error bars. The sample with thinnest MgO top layer ($t_{MgO}=4$~ML) however shows further enhanced dependence on the annealing temperature, which is shown in Fig.\ref{fig:PES2}. For $T_A$~=~400°C the spectrum is that of an insulator with a complete absence of photoemission yield down to 1.5~eV below $E_F$. We explain this as follows: Despite the annealing procedure the presence of very high defect density at the MgO surface, which at such very low MgO thickness is strongly correlated with the  defect density in the MgO film as well as at the CMS/MgO interface, leads to  highly augmented inelastic electron scattering. Due to this fact we effectively probe only the MgO layer and not the CMS/MgO interface. Elevating the annealing temperature to 500°C, a huge workfunction increase of 1.3~eV occurs as well as a finite photoemission yield near $E_F$, which now allows to determine the spin polarization at the Fermi energy. Obviously the annealing process induces defect healing, improving the interface and surface properties.  Higher annealing to the optimum value of 600°C only results in a further increase of $\Phi$ by 0.45~eV, while the SP does not change anymore. Here we want to recall that no LEED pattern could obtained for this sample. \newline 
To conclude, such drastic changes regarding the photoemission spectra do not appear at the other samples with increased MgO layer thickness, since the surface is less affected by interface-induced defects. This is in accordance with our AES and LEED results.

%%%%%%%%%%%%%%%%%%%%%%%%%%%%%%%%%%%%%%%%%%%%%%%%%%%%%%%%%%%%%%%%%

\section{Summary}
\label{sec:sum}

We have studied the formation of the $\mathrm{Co_{2}MnSi}$/MgO interface 
by considering $\mathrm{Co_{2}MnSi}$/MgO samples with variable MgO top layer thickness and annealing temperature. Distinct influence of interface defects onto the whole MgO layer is revealed by Auger electron spectroscopy and LEED for MgO thicknesses up to 10~ML, which can only partly be overcome by annealing. Low-energy SR-PES was used to determine the  spin polarization of the CMS/MgO interface buried below the MgO layers with thickness up to 20~ML, finding distinct positive spin-polarization values in all cases. The photoemission results obtained for different annealing temperatures  fully agree with the findings made by AES and LEED.

%%%%%%%%%%%%%%%%%%%%%%%%%%%%%%%%%%%%%%%%%%%%%%%%%%%%%%%%%%%%%%%%%
\section{Acknowledgement}
The authors want to gratefully acknowledge financial support through the DFG Research Unit 1464 ASPIMATT. 
The work at Hokkaido University was partly supported by a Grantin-Aid for Scientific Research (A) (Grant No. 23246055) from MEXT, Japan.

%%%%%%%%%%%%%%%%%%%%%%%%%%%%%%%%%%%%%%%%%%%%%%%%%%%%%%%%%%%%%%%%%

%\bibliographystyle{unsrt}
%\bibliography{Fetzer}

%merlin.mbs 2010-03-15 4.21a (PWD, AO, DPC)
\begin{thebibliography}{69}%
\makeatletter
\providecommand \@ifxundefined [1]{%
 \@ifx{#1\undefined}
}%
\providecommand \@ifnum [1]{%
 \ifnum #1\expandafter \@firstoftwo
 \else \expandafter \@secondoftwo
 \fi
}%
\providecommand \@ifx [1]{%
 \ifx #1\expandafter \@firstoftwo
 \else \expandafter \@secondoftwo
 \fi
}%
\providecommand \natexlab [1]{#1}%
\providecommand \enquote  [1]{``#1''}%
\providecommand \bibnamefont  [1]{#1}%
\providecommand \bibfnamefont [1]{#1}%
\providecommand \citenamefont [1]{#1}%
\providecommand \href@noop [0]{\@secondoftwo}%
\providecommand \href [0]{\begingroup \@sanitize@url \@href}%
\providecommand \@href[1]{\@@startlink{#1}\@@href}%
\providecommand \@@href[1]{\endgroup#1\@@endlink}%
\providecommand \@sanitize@url [0]{\catcode `\\12\catcode `\$12\catcode
  `\&12\catcode `\#12\catcode `\^12\catcode `\_12\catcode `\%12\relax}%
\providecommand \@@startlink[1]{}%
\providecommand \@@endlink[0]{}%
\providecommand \url  [0]{\begingroup\@sanitize@url \@url }%
\providecommand \@url [1]{\endgroup\@href {#1}{\urlprefix }}%
\providecommand \urlprefix  [0]{URL }%
\providecommand \Eprint [0]{\href }%
\@ifxundefined \urlstyle {%
  \providecommand \doi  [0]{\begingroup \@sanitize@url \@doi}%
  \providecommand \@doi [1]{\endgroup \@@startlink {\doibase
  #1}doi:\discretionary {}{}{}#1\@@endlink }%
}{%
  \providecommand \doi  [0]{doi:\discretionary{}{}{}\begingroup
  \urlstyle{rm}\Url }%
}%
\providecommand \doibase [0]{http://dx.doi.org/}%
\providecommand \Doi [0]{\begingroup \@sanitize@url \@Doi }%
\providecommand \@Doi  [1]{\endgroup\@@startlink{\doibase#1}\@@Doi}%
\providecommand \@@Doi [1]{#1\@@endlink}%
\providecommand \selectlanguage [0]{\@gobble}%
\providecommand \bibinfo  [0]{\@secondoftwo}%
\providecommand \bibfield  [0]{\@secondoftwo}%
\providecommand \translation [1]{[#1]}%
\providecommand \BibitemOpen [0]{}%
\providecommand \bibitemStop [0]{}%
\providecommand \bibitemNoStop [0]{.\EOS\space}%
\providecommand \EOS [0]{\spacefactor3000\relax}%
\providecommand \BibitemShut  [1]{\csname bibitem#1\endcsname}%
%</preamble>
\bibitem [{\citenamefont {Moodera}\ \emph {et~al.}(1995)\citenamefont
  {Moodera}, \citenamefont {Kinder}, \citenamefont {Wong},\ and\ \citenamefont
  {Meservey}}]{Moodera95}%
  \BibitemOpen
  \bibfield  {author} {\bibinfo {author} {\bibfnamefont {J.~S.}\ \bibnamefont
  {Moodera}}, \bibinfo {author} {\bibfnamefont {L.~R.}\ \bibnamefont {Kinder}},
  \bibinfo {author} {\bibfnamefont {T.~M.}\ \bibnamefont {Wong}}, \ and\
  \bibinfo {author} {\bibfnamefont {R.}~\bibnamefont {Meservey}},\ }\Doi
  {10.1103/PhysRevLett.74.3273} {\bibfield  {journal} {\bibinfo  {journal}
  {Phys. Rev. Lett.},\ }\textbf {\bibinfo {volume} {74}},\ \bibinfo {pages}
  {3273} (\bibinfo {year} {1995})}\BibitemShut {NoStop}%
\bibitem [{\citenamefont {Tsymbal}\ \emph {et~al.}(2003)\citenamefont
  {Tsymbal}, \citenamefont {Mryasov},\ and\ \citenamefont
  {LeClair}}]{Tsymbal03}%
  \BibitemOpen
  \bibfield  {author} {\bibinfo {author} {\bibfnamefont {E.~Y.}\ \bibnamefont
  {Tsymbal}}, \bibinfo {author} {\bibfnamefont {O.~N.}\ \bibnamefont
  {Mryasov}}, \ and\ \bibinfo {author} {\bibfnamefont {P.~R.}\ \bibnamefont
  {LeClair}},\ }\href {http://stacks.iop.org/0953-8984/15/i=4/a=201} {\bibfield
   {journal} {\bibinfo  {journal} {Journal of Physics: Condensed Matter},\
  }\textbf {\bibinfo {volume} {15}},\ \bibinfo {pages} {R109} (\bibinfo {year}
  {2003})}\BibitemShut {NoStop}%
\bibitem [{\citenamefont {Zutic}(2004)}]{Zutic04}%
  \BibitemOpen
  \bibfield  {author} {\bibinfo {author} {\bibfnamefont {I.}~\bibnamefont
  {Zutic}},\ }\href@noop {} {\bibfield  {journal} {\bibinfo  {journal} {Review
  Of Modern Physics},\ }\textbf {\bibinfo {volume} {76}} (\bibinfo {year}
  {2004})}\BibitemShut {NoStop}%
\bibitem [{\citenamefont {Velev}\ \emph {et~al.}(2008)\citenamefont {Velev},
  \citenamefont {Dowben}, \citenamefont {Tsymbal}, \citenamefont {Jenkins},\
  and\ \citenamefont {Caruso}}]{Velev08}%
  \BibitemOpen
  \bibfield  {author} {\bibinfo {author} {\bibfnamefont {J.}~\bibnamefont
  {Velev}}, \bibinfo {author} {\bibfnamefont {P.}~\bibnamefont {Dowben}},
  \bibinfo {author} {\bibfnamefont {E.}~\bibnamefont {Tsymbal}}, \bibinfo
  {author} {\bibfnamefont {S.}~\bibnamefont {Jenkins}}, \ and\ \bibinfo
  {author} {\bibfnamefont {A.}~\bibnamefont {Caruso}},\ }\Doi
  {10.1016/j.surfrep.2008.06.002} {\bibfield  {journal} {\bibinfo  {journal}
  {Surface Science Reports},\ }\textbf {\bibinfo {volume} {63}},\ \bibinfo
  {pages} {400 } (\bibinfo {year} {2008})},\ ISSN \bibinfo {issn}
  {0167-5729}\BibitemShut {NoStop}%
\bibitem [{\citenamefont {Julliere}(1975)}]{Julliere75}%
  \BibitemOpen
  \bibfield  {author} {\bibinfo {author} {\bibfnamefont {M.}~\bibnamefont
  {Julliere}},\ }\Doi {DOI: 10.1016/0375-9601(75)90174-7} {\bibfield  {journal}
  {\bibinfo  {journal} {Phys. Lett. A},\ }\textbf {\bibinfo {volume} {54}},\
  \bibinfo {pages} {225 } (\bibinfo {year} {1975})},\ ISSN \bibinfo {issn}
  {0375-9601}\BibitemShut {NoStop}%
\bibitem [{\citenamefont {Brown}\ \emph {et~al.}(2000)\citenamefont {Brown},
  \citenamefont {Neumann}, \citenamefont {Webster},\ and\ \citenamefont
  {Ziebeck}}]{Brown00}%
  \BibitemOpen
  \bibfield  {author} {\bibinfo {author} {\bibfnamefont {P.~J.}\ \bibnamefont
  {Brown}}, \bibinfo {author} {\bibfnamefont {K.~U.}\ \bibnamefont {Neumann}},
  \bibinfo {author} {\bibfnamefont {P.~J.}\ \bibnamefont {Webster}}, \ and\
  \bibinfo {author} {\bibfnamefont {K.~R.~A.}\ \bibnamefont {Ziebeck}},\ }\href
  {http://stacks.iop.org/0953-8984/12/1827} {\bibfield  {journal} {\bibinfo
  {journal} {Journal of Physics: Condensed Matter},\ }\textbf {\bibinfo
  {volume} {12}},\ \bibinfo {pages} {1827} (\bibinfo {year}
  {2000})}\BibitemShut {NoStop}%
\bibitem [{\citenamefont {K\"ammerer}\ \emph {et~al.}(2004)\citenamefont
  {K\"ammerer}, \citenamefont {Thomas}, \citenamefont {Hutten},\ and\
  \citenamefont {Reiss}}]{Kaemmerer04}%
  \BibitemOpen
  \bibfield  {author} {\bibinfo {author} {\bibfnamefont {S.}~\bibnamefont
  {K\"ammerer}}, \bibinfo {author} {\bibfnamefont {A.}~\bibnamefont {Thomas}},
  \bibinfo {author} {\bibfnamefont {A.}~\bibnamefont {Hutten}}, \ and\ \bibinfo
  {author} {\bibfnamefont {G.}~\bibnamefont {Reiss}},\ }\href
  {http://link.aip.org/link/?APL/85/79/1} {\bibfield  {journal} {\bibinfo
  {journal} {Appl. Phys. Lett.},\ }\textbf {\bibinfo {volume} {85}},\ \bibinfo
  {pages} {79} (\bibinfo {year} {2004})}\BibitemShut {NoStop}%
\bibitem [{\citenamefont {Ishikawa}\ \emph {et~al.}(2006)\citenamefont
  {Ishikawa}, \citenamefont {Marukame}, \citenamefont {Kijima}, \citenamefont
  {Matsuda}, \citenamefont {Uemura}, \citenamefont {Arita},\ and\ \citenamefont
  {Yamamoto}}]{Ishikawa06}%
  \BibitemOpen
  \bibfield  {author} {\bibinfo {author} {\bibfnamefont {T.}~\bibnamefont
  {Ishikawa}}, \bibinfo {author} {\bibfnamefont {T.}~\bibnamefont {Marukame}},
  \bibinfo {author} {\bibfnamefont {H.}~\bibnamefont {Kijima}}, \bibinfo
  {author} {\bibfnamefont {K.~I.}\ \bibnamefont {Matsuda}}, \bibinfo {author}
  {\bibfnamefont {T.}~\bibnamefont {Uemura}}, \bibinfo {author} {\bibfnamefont
  {M.}~\bibnamefont {Arita}}, \ and\ \bibinfo {author} {\bibfnamefont
  {M.}~\bibnamefont {Yamamoto}},\ }\Doi {10.1063/1.2378397} {\bibfield
  {journal} {\bibinfo  {journal} {Appl. Phys. Lett.},\ }\textbf {\bibinfo
  {volume} {89}} (\bibinfo {year} {2006})},\ \doi
  {10.1063/1.2378397}\BibitemShut {NoStop}%
\bibitem [{\citenamefont {Ishida}\ \emph {et~al.}(1998)\citenamefont {Ishida},
  \citenamefont {Masaki}, \citenamefont {Fujii},\ and\ \citenamefont
  {Asano}}]{Ishida98}%
  \BibitemOpen
  \bibfield  {author} {\bibinfo {author} {\bibfnamefont {S.}~\bibnamefont
  {Ishida}}, \bibinfo {author} {\bibfnamefont {T.}~\bibnamefont {Masaki}},
  \bibinfo {author} {\bibfnamefont {S.}~\bibnamefont {Fujii}}, \ and\ \bibinfo
  {author} {\bibfnamefont {S.}~\bibnamefont {Asano}},\ }\Doi {DOI:
  10.1016/S0921-4526(97)00495-X} {\bibfield  {journal} {\bibinfo  {journal}
  {Physica B: Condensed Matter},\ }\textbf {\bibinfo {volume} {245}},\ \bibinfo
  {pages} {1 } (\bibinfo {year} {1998})},\ ISSN \bibinfo {issn}
  {0921-4526}\BibitemShut {NoStop}%
\bibitem [{\citenamefont {Balke}\ \emph {et~al.}(2006)\citenamefont {Balke},
  \citenamefont {Fecher}, \citenamefont {Kandpal}, \citenamefont {Felser},
  \citenamefont {Kobayashi}, \citenamefont {Ikenaga}, \citenamefont {Kim},\
  and\ \citenamefont {Ueda}}]{Balke06}%
  \BibitemOpen
  \bibfield  {author} {\bibinfo {author} {\bibfnamefont {B.}~\bibnamefont
  {Balke}}, \bibinfo {author} {\bibfnamefont {G.~H.}\ \bibnamefont {Fecher}},
  \bibinfo {author} {\bibfnamefont {H.~C.}\ \bibnamefont {Kandpal}}, \bibinfo
  {author} {\bibfnamefont {C.}~\bibnamefont {Felser}}, \bibinfo {author}
  {\bibfnamefont {K.}~\bibnamefont {Kobayashi}}, \bibinfo {author}
  {\bibfnamefont {E.}~\bibnamefont {Ikenaga}}, \bibinfo {author} {\bibfnamefont
  {J.-J.}\ \bibnamefont {Kim}}, \ and\ \bibinfo {author} {\bibfnamefont
  {S.}~\bibnamefont {Ueda}},\ }\Doi {10.1103/PhysRevB.74.104405} {\bibfield
  {journal} {\bibinfo  {journal} {Physical Review B (Condensed Matter and
  Materials Physics)},\ }\textbf {\bibinfo {volume} {74}},\ \bibinfo {eid}
  {104405} (\bibinfo {year} {2006})}\BibitemShut {NoStop}%
\bibitem [{\citenamefont {Chadov}\ \emph {et~al.}(2009)\citenamefont {Chadov},
  \citenamefont {Fecher}, \citenamefont {Felser}, \citenamefont {Minar},
  \citenamefont {Braun},\ and\ \citenamefont {Ebert}}]{Chadov09}%
  \BibitemOpen
  \bibfield  {author} {\bibinfo {author} {\bibfnamefont {S.}~\bibnamefont
  {Chadov}}, \bibinfo {author} {\bibfnamefont {G.~H.}\ \bibnamefont {Fecher}},
  \bibinfo {author} {\bibfnamefont {C.}~\bibnamefont {Felser}}, \bibinfo
  {author} {\bibfnamefont {J.}~\bibnamefont {Minar}}, \bibinfo {author}
  {\bibfnamefont {J.}~\bibnamefont {Braun}}, \ and\ \bibinfo {author}
  {\bibfnamefont {H.}~\bibnamefont {Ebert}},\ }\href
  {http://stacks.iop.org/0022-3727/42/i=8/a=084002} {\bibfield  {journal}
  {\bibinfo  {journal} {Journal of Physics D: Applied Physics},\ }\textbf
  {\bibinfo {volume} {42}},\ \bibinfo {pages} {084002} (\bibinfo {year}
  {2009})}\BibitemShut {NoStop}%
\bibitem [{\citenamefont {Sakuraba}\ \emph
  {et~al.}(2006){\natexlab{a}}\citenamefont {Sakuraba}, \citenamefont
  {Miyakoshi}, \citenamefont {Oogane}, \citenamefont {Ando}, \citenamefont
  {Sakuma}, \citenamefont {Miyazaki},\ and\ \citenamefont
  {Kubota}}]{Sakuraba06b}%
  \BibitemOpen
  \bibfield  {author} {\bibinfo {author} {\bibfnamefont {Y.}~\bibnamefont
  {Sakuraba}}, \bibinfo {author} {\bibfnamefont {T.}~\bibnamefont {Miyakoshi}},
  \bibinfo {author} {\bibfnamefont {M.}~\bibnamefont {Oogane}}, \bibinfo
  {author} {\bibfnamefont {Y.}~\bibnamefont {Ando}}, \bibinfo {author}
  {\bibfnamefont {A.}~\bibnamefont {Sakuma}}, \bibinfo {author} {\bibfnamefont
  {T.}~\bibnamefont {Miyazaki}}, \ and\ \bibinfo {author} {\bibfnamefont
  {H.}~\bibnamefont {Kubota}},\ }\Doi {10.1063/1.2335583} {\bibfield  {journal}
  {\bibinfo  {journal} {Applied Physics Letters},\ }\textbf {\bibinfo {volume}
  {89}},\ \bibinfo {eid} {052508} (\bibinfo {year}
  {2006}{\natexlab{a}})}\BibitemShut {NoStop}%
\bibitem [{\citenamefont {Ishikawa}\ \emph
  {et~al.}(2009){\natexlab{a}}\citenamefont {Ishikawa}, \citenamefont
  {Itabashi}, \citenamefont {Taira}, \citenamefont {ichi Matsuda},
  \citenamefont {Uemura},\ and\ \citenamefont {Yamamoto}}]{Ishikawa09a}%
  \BibitemOpen
  \bibfield  {author} {\bibinfo {author} {\bibfnamefont {T.}~\bibnamefont
  {Ishikawa}}, \bibinfo {author} {\bibfnamefont {N.}~\bibnamefont {Itabashi}},
  \bibinfo {author} {\bibfnamefont {T.}~\bibnamefont {Taira}}, \bibinfo
  {author} {\bibfnamefont {K.}~\bibnamefont {ichi Matsuda}}, \bibinfo {author}
  {\bibfnamefont {T.}~\bibnamefont {Uemura}}, \ and\ \bibinfo {author}
  {\bibfnamefont {M.}~\bibnamefont {Yamamoto}},\ }\Doi {10.1063/1.3083560}
  {\bibfield  {journal} {\bibinfo  {journal} {Applied Physics Letters},\
  }\textbf {\bibinfo {volume} {94}},\ \bibinfo {eid} {092503} (\bibinfo {year}
  {2009}{\natexlab{a}})}\BibitemShut {NoStop}%
\bibitem [{\citenamefont {Sakuraba}\ \emph
  {et~al.}(2006){\natexlab{b}}\citenamefont {Sakuraba}, \citenamefont
  {Hattori}, \citenamefont {Oogane}, \citenamefont {Ando}, \citenamefont
  {Kato}, \citenamefont {Sakuma}, \citenamefont {Miyazaki},\ and\ \citenamefont
  {Kubota}}]{Sakuraba06a}%
  \BibitemOpen
  \bibfield  {author} {\bibinfo {author} {\bibfnamefont {Y.}~\bibnamefont
  {Sakuraba}}, \bibinfo {author} {\bibfnamefont {M.}~\bibnamefont {Hattori}},
  \bibinfo {author} {\bibfnamefont {M.}~\bibnamefont {Oogane}}, \bibinfo
  {author} {\bibfnamefont {Y.}~\bibnamefont {Ando}}, \bibinfo {author}
  {\bibfnamefont {H.}~\bibnamefont {Kato}}, \bibinfo {author} {\bibfnamefont
  {A.}~\bibnamefont {Sakuma}}, \bibinfo {author} {\bibfnamefont
  {T.}~\bibnamefont {Miyazaki}}, \ and\ \bibinfo {author} {\bibfnamefont
  {H.}~\bibnamefont {Kubota}},\ }\Doi {10.1063/1.2202724} {\bibfield  {journal}
  {\bibinfo  {journal} {Applied Physics Letters},\ }\textbf {\bibinfo {volume}
  {88}},\ \bibinfo {eid} {192508} (\bibinfo {year}
  {2006}{\natexlab{b}})}\BibitemShut {NoStop}%
\bibitem [{\citenamefont {Ishikawa}\ \emph {et~al.}(2008)\citenamefont
  {Ishikawa}, \citenamefont {Hakamata}, \citenamefont {Matsuda}, \citenamefont
  {Uemura},\ and\ \citenamefont {Yamamoto}}]{Ishikawa08}%
  \BibitemOpen
  \bibfield  {author} {\bibinfo {author} {\bibfnamefont {T.}~\bibnamefont
  {Ishikawa}}, \bibinfo {author} {\bibfnamefont {S.}~\bibnamefont {Hakamata}},
  \bibinfo {author} {\bibfnamefont {K.}~\bibnamefont {Matsuda}}, \bibinfo
  {author} {\bibfnamefont {T.}~\bibnamefont {Uemura}}, \ and\ \bibinfo {author}
  {\bibfnamefont {M.}~\bibnamefont {Yamamoto}},\ }\Doi {10.1063/1.2843756}
  {\bibfield  {journal} {\bibinfo  {journal} {J. Appl. Phys.},\ }\textbf
  {\bibinfo {volume} {103}},\ \bibinfo {pages} {07A919} (\bibinfo {year}
  {2008})}\BibitemShut {NoStop}%
\bibitem [{\citenamefont {Butler}\ \emph {et~al.}(2001)\citenamefont {Butler},
  \citenamefont {Zhang}, \citenamefont {Schulthess},\ and\ \citenamefont
  {MacLaren}}]{Butler01}%
  \BibitemOpen
  \bibfield  {author} {\bibinfo {author} {\bibfnamefont {W.~H.}\ \bibnamefont
  {Butler}}, \bibinfo {author} {\bibfnamefont {X.-G.}\ \bibnamefont {Zhang}},
  \bibinfo {author} {\bibfnamefont {T.~C.}\ \bibnamefont {Schulthess}}, \ and\
  \bibinfo {author} {\bibfnamefont {J.~M.}\ \bibnamefont {MacLaren}},\ }\Doi
  {10.1103/PhysRevB.63.054416} {\bibfield  {journal} {\bibinfo  {journal}
  {Phys. Rev. B},\ }\textbf {\bibinfo {volume} {63}},\ \bibinfo {pages}
  {054416} (\bibinfo {year} {2001})}\BibitemShut {NoStop}%
\bibitem [{\citenamefont {Miura}\ \emph {et~al.}(2007)\citenamefont {Miura},
  \citenamefont {Uchida}, \citenamefont {Oba}, \citenamefont {Nagao},\ and\
  \citenamefont {Shirai}}]{Miura07}%
  \BibitemOpen
  \bibfield  {author} {\bibinfo {author} {\bibfnamefont {Y.}~\bibnamefont
  {Miura}}, \bibinfo {author} {\bibfnamefont {H.}~\bibnamefont {Uchida}},
  \bibinfo {author} {\bibfnamefont {Y.}~\bibnamefont {Oba}}, \bibinfo {author}
  {\bibfnamefont {K.}~\bibnamefont {Nagao}}, \ and\ \bibinfo {author}
  {\bibfnamefont {M.}~\bibnamefont {Shirai}},\ }\Doi
  {10.1088/0953-8984/19/36/365228} {\bibfield  {journal} {\bibinfo  {journal}
  {J. Phys.: Condens. Matter},\ }\textbf {\bibinfo {volume} {19}} (\bibinfo
  {year} {2007})},\ \doi {10.1088/0953-8984/19/36/365228}\BibitemShut {NoStop}%
\bibitem [{\citenamefont {Ishikawa}\ \emph
  {et~al.}(2009){\natexlab{b}}\citenamefont {Ishikawa}, \citenamefont {Liu},
  \citenamefont {Taira}, \citenamefont {Matsuda}, \citenamefont {Uemura},\ and\
  \citenamefont {Yamamoto}}]{Ishikawa09b}%
  \BibitemOpen
  \bibfield  {author} {\bibinfo {author} {\bibfnamefont {T.}~\bibnamefont
  {Ishikawa}}, \bibinfo {author} {\bibfnamefont {H.}~\bibnamefont {Liu}},
  \bibinfo {author} {\bibfnamefont {T.}~\bibnamefont {Taira}}, \bibinfo
  {author} {\bibfnamefont {K.}~\bibnamefont {Matsuda}}, \bibinfo {author}
  {\bibfnamefont {T.}~\bibnamefont {Uemura}}, \ and\ \bibinfo {author}
  {\bibfnamefont {M.}~\bibnamefont {Yamamoto}},\ }\href@noop {} {\bibfield
  {journal} {\bibinfo  {journal} {Applied Physics Letters},\ }\textbf {\bibinfo
  {volume} {95}},\ \bibinfo {pages} {232512} (\bibinfo {year}
  {2009}{\natexlab{b}})}\BibitemShut {NoStop}%
\bibitem [{\citenamefont {Chang}\ and\ \citenamefont {Cohen}(1984)}]{Chang84}%
  \BibitemOpen
  \bibfield  {author} {\bibinfo {author} {\bibfnamefont {K.~J.}\ \bibnamefont
  {Chang}}\ and\ \bibinfo {author} {\bibfnamefont {M.~L.}\ \bibnamefont
  {Cohen}},\ }\Doi {10.1103/PhysRevB.30.4774} {\bibfield  {journal} {\bibinfo
  {journal} {Phys. Rev. B},\ }\textbf {\bibinfo {volume} {30}},\ \bibinfo
  {pages} {4774} (\bibinfo {year} {1984})}\BibitemShut {NoStop}%
\bibitem [{\citenamefont {Xu}\ and\ \citenamefont {Ching}(1991)}]{Xu91}%
  \BibitemOpen
  \bibfield  {author} {\bibinfo {author} {\bibfnamefont {Y.-N.}\ \bibnamefont
  {Xu}}\ and\ \bibinfo {author} {\bibfnamefont {W.~Y.}\ \bibnamefont {Ching}},\
  }\Doi {10.1103/PhysRevB.43.4461} {\bibfield  {journal} {\bibinfo  {journal}
  {Phys. Rev. B},\ }\textbf {\bibinfo {volume} {43}},\ \bibinfo {pages} {4461}
  (\bibinfo {year} {1991})}\BibitemShut {NoStop}%
\bibitem [{\citenamefont {Roessler}\ and\ \citenamefont
  {Walker}(1967)}]{Roessler67}%
  \BibitemOpen
  \bibfield  {author} {\bibinfo {author} {\bibfnamefont {D.~M.}\ \bibnamefont
  {Roessler}}\ and\ \bibinfo {author} {\bibfnamefont {W.~C.}\ \bibnamefont
  {Walker}},\ }\Doi {10.1103/PhysRev.159.733} {\bibfield  {journal} {\bibinfo
  {journal} {Phys. Rev.},\ }\textbf {\bibinfo {volume} {159}},\ \bibinfo
  {pages} {733} (\bibinfo {year} {1967})}\BibitemShut {NoStop}%
\bibitem [{\citenamefont {Pandey}\ \emph {et~al.}(1991)\citenamefont {Pandey},
  \citenamefont {Jaffe},\ and\ \citenamefont {Kunz}}]{Pandey91}%
  \BibitemOpen
  \bibfield  {author} {\bibinfo {author} {\bibfnamefont {R.}~\bibnamefont
  {Pandey}}, \bibinfo {author} {\bibfnamefont {J.~E.}\ \bibnamefont {Jaffe}}, \
  and\ \bibinfo {author} {\bibfnamefont {A.~B.}\ \bibnamefont {Kunz}},\ }\Doi
  {10.1103/PhysRevB.43.9228} {\bibfield  {journal} {\bibinfo  {journal} {Phys.
  Rev. B},\ }\textbf {\bibinfo {volume} {43}},\ \bibinfo {pages} {9228}
  (\bibinfo {year} {1991})}\BibitemShut {NoStop}%
\bibitem [{\citenamefont {Gibson}\ \emph {et~al.}(1994)\citenamefont {Gibson},
  \citenamefont {Haydock},\ and\ \citenamefont {LaFemina}}]{Gibson94}%
  \BibitemOpen
  \bibfield  {author} {\bibinfo {author} {\bibfnamefont {A.}~\bibnamefont
  {Gibson}}, \bibinfo {author} {\bibfnamefont {R.}~\bibnamefont {Haydock}}, \
  and\ \bibinfo {author} {\bibfnamefont {J.~P.}\ \bibnamefont {LaFemina}},\
  }\Doi {10.1103/PhysRevB.50.2582} {\bibfield  {journal} {\bibinfo  {journal}
  {Phys. Rev. B},\ }\textbf {\bibinfo {volume} {50}},\ \bibinfo {pages} {2582}
  (\bibinfo {year} {1994})}\BibitemShut {NoStop}%
\bibitem [{\citenamefont {Kantorovich}\ \emph {et~al.}(1995)\citenamefont
  {Kantorovich}, \citenamefont {Holender},\ and\ \citenamefont
  {Gillan}}]{Kantorovich95}%
  \BibitemOpen
  \bibfield  {author} {\bibinfo {author} {\bibfnamefont {L.}~\bibnamefont
  {Kantorovich}}, \bibinfo {author} {\bibfnamefont {J.}~\bibnamefont
  {Holender}}, \ and\ \bibinfo {author} {\bibfnamefont {M.}~\bibnamefont
  {Gillan}},\ }\href
  {http://www.sciencedirect.com/science/article/pii/0039602895008446}
  {\bibfield  {journal} {\bibinfo  {journal} {Surface Science},\ }\textbf
  {\bibinfo {volume} {343}},\ \bibinfo {pages} {221} (\bibinfo {year}
  {1995})},\ ISSN \bibinfo {issn} {0039-6028}\BibitemShut {NoStop}%
\bibitem [{\citenamefont {Illas}\ and\ \citenamefont
  {Pacchioni}(1998)}]{Illas98}%
  \BibitemOpen
  \bibfield  {author} {\bibinfo {author} {\bibfnamefont {F.}~\bibnamefont
  {Illas}}\ and\ \bibinfo {author} {\bibfnamefont {G.}~\bibnamefont
  {Pacchioni}},\ }\Doi {10.1063/1.476220} {\bibfield  {journal} {\bibinfo
  {journal} {The Journal of Chemical Physics},\ }\textbf {\bibinfo {volume}
  {108}},\ \bibinfo {pages} {7835} (\bibinfo {year} {1998})}\BibitemShut
  {NoStop}%
\bibitem [{\citenamefont {Lee}\ and\ \citenamefont {Wong}(1978)}]{Lee78}%
  \BibitemOpen
  \bibfield  {author} {\bibinfo {author} {\bibfnamefont {V.-C.}\ \bibnamefont
  {Lee}}\ and\ \bibinfo {author} {\bibfnamefont {H.-S.}\ \bibnamefont {Wong}},\
  }\Doi {10.1143/JPSJ.45.895} {\bibfield  {journal} {\bibinfo  {journal}
  {Journal of the Physical Society of Japan},\ }\textbf {\bibinfo {volume}
  {45}},\ \bibinfo {pages} {895} (\bibinfo {year} {1978})}\BibitemShut
  {NoStop}%
\bibitem [{\citenamefont {Sch\"onberger}\ and\ \citenamefont
  {Aryasetiawan}(1995)}]{Schoenberger95}%
  \BibitemOpen
  \bibfield  {author} {\bibinfo {author} {\bibfnamefont {U.}~\bibnamefont
  {Sch\"onberger}}\ and\ \bibinfo {author} {\bibfnamefont {F.}~\bibnamefont
  {Aryasetiawan}},\ }\Doi {10.1103/PhysRevB.52.8788} {\bibfield  {journal}
  {\bibinfo  {journal} {Phys. Rev. B},\ }\textbf {\bibinfo {volume} {52}},\
  \bibinfo {pages} {8788} (\bibinfo {year} {1995})}\BibitemShut {NoStop}%
\bibitem [{\citenamefont {Mather}\ \emph {et~al.}(2006)\citenamefont {Mather},
  \citenamefont {Read},\ and\ \citenamefont {Buhrman}}]{Mather06}%
  \BibitemOpen
  \bibfield  {author} {\bibinfo {author} {\bibfnamefont {P.~G.}\ \bibnamefont
  {Mather}}, \bibinfo {author} {\bibfnamefont {J.~C.}\ \bibnamefont {Read}}, \
  and\ \bibinfo {author} {\bibfnamefont {R.~A.}\ \bibnamefont {Buhrman}},\
  }\Doi {10.1103/PhysRevB.73.205412} {\bibfield  {journal} {\bibinfo  {journal}
  {Phys. Rev. B},\ }\textbf {\bibinfo {volume} {73}},\ \bibinfo {pages}
  {205412} (\bibinfo {year} {2006})}\BibitemShut {NoStop}%
\bibitem [{\citenamefont {Sterrer}\ \emph {et~al.}(2006)\citenamefont
  {Sterrer}, \citenamefont {Fischbach}, \citenamefont {Heyde}, \citenamefont
  {Nilius}, \citenamefont {Rust}, \citenamefont {Risse},\ and\ \citenamefont
  {Freund}}]{Sterrer06b}%
  \BibitemOpen
  \bibfield  {author} {\bibinfo {author} {\bibfnamefont {M.}~\bibnamefont
  {Sterrer}}, \bibinfo {author} {\bibfnamefont {E.}~\bibnamefont {Fischbach}},
  \bibinfo {author} {\bibfnamefont {M.}~\bibnamefont {Heyde}}, \bibinfo
  {author} {\bibfnamefont {N.}~\bibnamefont {Nilius}}, \bibinfo {author}
  {\bibfnamefont {H.-P.}\ \bibnamefont {Rust}}, \bibinfo {author}
  {\bibfnamefont {T.}~\bibnamefont {Risse}}, \ and\ \bibinfo {author}
  {\bibfnamefont {H.-J.}\ \bibnamefont {Freund}},\ }\Doi {10.1021/jp060546t}
  {\bibfield  {journal} {\bibinfo  {journal} {The Journal of Physical Chemistry
  B},\ }\textbf {\bibinfo {volume} {110}},\ \bibinfo {pages} {8665} (\bibinfo
  {year} {2006})},\ \Eprint
  {http://arxiv.org/abs/http://pubs.acs.org/doi/pdf/10.1021/jp060546t}
  {http://pubs.acs.org/doi/pdf/10.1021/jp060546t} \BibitemShut {NoStop}%
\bibitem [{\citenamefont {Afanas'ev}\ \emph {et~al.}(2010)\citenamefont
  {Afanas'ev}, \citenamefont {Stesmans}, \citenamefont {Cherkaoui},\ and\
  \citenamefont {Hurley}}]{Afanasev10}%
  \BibitemOpen
  \bibfield  {author} {\bibinfo {author} {\bibfnamefont {V.~V.}\ \bibnamefont
  {Afanas'ev}}, \bibinfo {author} {\bibfnamefont {A.}~\bibnamefont {Stesmans}},
  \bibinfo {author} {\bibfnamefont {K.}~\bibnamefont {Cherkaoui}}, \ and\
  \bibinfo {author} {\bibfnamefont {P.~K.}\ \bibnamefont {Hurley}},\ }\Doi
  {10.1063/1.3294328} {\bibfield  {journal} {\bibinfo  {journal} {Applied
  Physics Letters},\ }\textbf {\bibinfo {volume} {96}},\ \bibinfo {eid}
  {052103} (\bibinfo {year} {2010})}\BibitemShut {NoStop}%
\bibitem [{\citenamefont {Klaua}\ \emph {et~al.}(2001)\citenamefont {Klaua},
  \citenamefont {Ullmann}, \citenamefont {Barthel}, \citenamefont {Wulfhekel},
  \citenamefont {Kirschner}, \citenamefont {Urban}, \citenamefont {Monchesky},
  \citenamefont {Enders}, \citenamefont {Cochran},\ and\ \citenamefont
  {Heinrich}}]{Klaua01}%
  \BibitemOpen
  \bibfield  {author} {\bibinfo {author} {\bibfnamefont {M.}~\bibnamefont
  {Klaua}}, \bibinfo {author} {\bibfnamefont {D.}~\bibnamefont {Ullmann}},
  \bibinfo {author} {\bibfnamefont {J.}~\bibnamefont {Barthel}}, \bibinfo
  {author} {\bibfnamefont {W.}~\bibnamefont {Wulfhekel}}, \bibinfo {author}
  {\bibfnamefont {J.}~\bibnamefont {Kirschner}}, \bibinfo {author}
  {\bibfnamefont {R.}~\bibnamefont {Urban}}, \bibinfo {author} {\bibfnamefont
  {T.~L.}\ \bibnamefont {Monchesky}}, \bibinfo {author} {\bibfnamefont
  {A.}~\bibnamefont {Enders}}, \bibinfo {author} {\bibfnamefont {J.~F.}\
  \bibnamefont {Cochran}}, \ and\ \bibinfo {author} {\bibfnamefont
  {B.}~\bibnamefont {Heinrich}},\ }\Doi {10.1103/PhysRevB.64.134411} {\bibfield
   {journal} {\bibinfo  {journal} {Phys. Rev. B},\ }\textbf {\bibinfo {volume}
  {64}},\ \bibinfo {pages} {134411} (\bibinfo {year} {2001})}\BibitemShut
  {NoStop}%
\bibitem [{\citenamefont {Dedkov}\ \emph {et~al.}(2006)\citenamefont {Dedkov},
  \citenamefont {Fonin}, \citenamefont {R\"udiger},\ and\ \citenamefont
  {G\"untherodt}}]{Dedkov06}%
  \BibitemOpen
  \bibfield  {author} {\bibinfo {author} {\bibfnamefont {Y.}~\bibnamefont
  {Dedkov}}, \bibinfo {author} {\bibfnamefont {M.}~\bibnamefont {Fonin}},
  \bibinfo {author} {\bibfnamefont {U.}~\bibnamefont {R\"udiger}}, \ and\
  \bibinfo {author} {\bibfnamefont {G.}~\bibnamefont {G\"untherodt}},\ }\href
  {http://dx.doi.org/10.1007/s00339-005-3447-2} {\bibfield  {journal} {\bibinfo
   {journal} {Applied Physics A: Materials Science \& Processing},\ }\textbf
  {\bibinfo {volume} {82}},\ \bibinfo {pages} {489} (\bibinfo {year} {2006})},\
  ISSN \bibinfo {issn} {0947-8396},\ \bibinfo {note}
  {10.1007/s00339-005-3447-2}\BibitemShut {NoStop}%
\bibitem [{\citenamefont {Miyajima}\ \emph {et~al.}(2009)\citenamefont
  {Miyajima}, \citenamefont {Oogane}, \citenamefont {Kotaka}, \citenamefont
  {Yamazaki}, \citenamefont {Tsukada}, \citenamefont {Kataoka}, \citenamefont
  {Naganuma},\ and\ \citenamefont {Ando}}]{Miyajima09}%
  \BibitemOpen
  \bibfield  {author} {\bibinfo {author} {\bibfnamefont {T.}~\bibnamefont
  {Miyajima}}, \bibinfo {author} {\bibfnamefont {M.}~\bibnamefont {Oogane}},
  \bibinfo {author} {\bibfnamefont {Y.}~\bibnamefont {Kotaka}}, \bibinfo
  {author} {\bibfnamefont {T.}~\bibnamefont {Yamazaki}}, \bibinfo {author}
  {\bibfnamefont {M.}~\bibnamefont {Tsukada}}, \bibinfo {author} {\bibfnamefont
  {Y.}~\bibnamefont {Kataoka}}, \bibinfo {author} {\bibfnamefont
  {H.}~\bibnamefont {Naganuma}}, \ and\ \bibinfo {author} {\bibfnamefont
  {Y.}~\bibnamefont {Ando}},\ }\Doi {10.1143/APEX.2.093001} {\bibfield
  {journal} {\bibinfo  {journal} {Appl. Phys. Express},\ }\textbf {\bibinfo
  {volume} {2}} (\bibinfo {year} {2009})},\ ISSN \bibinfo {issn}
  {{1882-0778}},\ \doi {10.1143/APEX.2.093001}\BibitemShut {NoStop}%
\bibitem [{\citenamefont {Saito}\ \emph {et~al.}(2007)\citenamefont {Saito},
  \citenamefont {Katayama}, \citenamefont {Ishikawa}, \citenamefont {Yamamoto},
  \citenamefont {Asakura},\ and\ \citenamefont {Koide}}]{Saito07}%
  \BibitemOpen
  \bibfield  {author} {\bibinfo {author} {\bibfnamefont {T.}~\bibnamefont
  {Saito}}, \bibinfo {author} {\bibfnamefont {T.}~\bibnamefont {Katayama}},
  \bibinfo {author} {\bibfnamefont {T.}~\bibnamefont {Ishikawa}}, \bibinfo
  {author} {\bibfnamefont {M.}~\bibnamefont {Yamamoto}}, \bibinfo {author}
  {\bibfnamefont {D.}~\bibnamefont {Asakura}}, \ and\ \bibinfo {author}
  {\bibfnamefont {T.}~\bibnamefont {Koide}},\ }\Doi {10.1063/1.2824856}
  {\bibfield  {journal} {\bibinfo  {journal} {Appl. Phys. Lett.},\ }\textbf
  {\bibinfo {volume} {91}} (\bibinfo {year} {2007})},\ \doi
  {10.1063/1.2824856}\BibitemShut {NoStop}%
\bibitem [{\citenamefont {Jourdan}\ \emph {et~al.}(2009)\citenamefont
  {Jourdan}, \citenamefont {Jorge}, \citenamefont {Herbort}, \citenamefont
  {Kallmayer}, \citenamefont {Klaer},\ and\ \citenamefont
  {Elmers}}]{Jourdan09}%
  \BibitemOpen
  \bibfield  {author} {\bibinfo {author} {\bibfnamefont {M.}~\bibnamefont
  {Jourdan}}, \bibinfo {author} {\bibfnamefont {E.~A.}\ \bibnamefont {Jorge}},
  \bibinfo {author} {\bibfnamefont {C.}~\bibnamefont {Herbort}}, \bibinfo
  {author} {\bibfnamefont {M.}~\bibnamefont {Kallmayer}}, \bibinfo {author}
  {\bibfnamefont {P.}~\bibnamefont {Klaer}}, \ and\ \bibinfo {author}
  {\bibfnamefont {H.-J.}\ \bibnamefont {Elmers}},\ }\Doi {10.1063/1.3254252}
  {\bibfield  {journal} {\bibinfo  {journal} {Applied Physics Letters},\
  }\textbf {\bibinfo {volume} {95}},\ \bibinfo {eid} {172504} (\bibinfo {year}
  {2009})}\BibitemShut {NoStop}%
\bibitem [{\citenamefont {Sakuraba}\ \emph {et~al.}(2010)\citenamefont
  {Sakuraba}, \citenamefont {Takanashi}, \citenamefont {Kota}, \citenamefont
  {Kubota}, \citenamefont {Oogane}, \citenamefont {Sakuma},\ and\ \citenamefont
  {Ando}}]{Sakuraba10}%
  \BibitemOpen
  \bibfield  {author} {\bibinfo {author} {\bibfnamefont {Y.}~\bibnamefont
  {Sakuraba}}, \bibinfo {author} {\bibfnamefont {K.}~\bibnamefont {Takanashi}},
  \bibinfo {author} {\bibfnamefont {Y.}~\bibnamefont {Kota}}, \bibinfo {author}
  {\bibfnamefont {T.}~\bibnamefont {Kubota}}, \bibinfo {author} {\bibfnamefont
  {M.}~\bibnamefont {Oogane}}, \bibinfo {author} {\bibfnamefont
  {A.}~\bibnamefont {Sakuma}}, \ and\ \bibinfo {author} {\bibfnamefont
  {Y.}~\bibnamefont {Ando}},\ }\Doi {10.1103/PhysRevB.81.144422} {\bibfield
  {journal} {\bibinfo  {journal} {Phys. Rev. B},\ }\textbf {\bibinfo {volume}
  {81}},\ \bibinfo {pages} {144422} (\bibinfo {year} {2010})}\BibitemShut
  {NoStop}%
\bibitem [{\citenamefont {Valeri}\ \emph {et~al.}(2001)\citenamefont {Valeri},
  \citenamefont {Altieri}, \citenamefont {di~Bona}, \citenamefont
  {Giovanardi},\ and\ \citenamefont {Moia}}]{Valeri01}%
  \BibitemOpen
  \bibfield  {author} {\bibinfo {author} {\bibfnamefont {S.}~\bibnamefont
  {Valeri}}, \bibinfo {author} {\bibfnamefont {S.}~\bibnamefont {Altieri}},
  \bibinfo {author} {\bibfnamefont {A.}~\bibnamefont {di~Bona}}, \bibinfo
  {author} {\bibfnamefont {C.}~\bibnamefont {Giovanardi}}, \ and\ \bibinfo
  {author} {\bibfnamefont {T.}~\bibnamefont {Moia}},\ }\bibfield  {booktitle}
  {\emph {\bibinfo {booktitle} {Proceedings of Symposium N on Ultrathin
  Oxides}},\ }\href
  {http://www.sciencedirect.com/science/article/pii/S0040609001014869}
  {\bibfield  {journal} {\bibinfo  {journal} {Thin Solid Films},\ }\textbf
  {\bibinfo {volume} {400}},\ \bibinfo {pages} {16} (\bibinfo {year} {2001})},\
  ISSN \bibinfo {issn} {0040-6090}\BibitemShut {NoStop}%
\bibitem [{\citenamefont {Vassent}\ \emph {et~al.}(1996)\citenamefont
  {Vassent}, \citenamefont {Dynna}, \citenamefont {Marty}, \citenamefont
  {Gilles},\ and\ \citenamefont {Patrat}}]{Vassent96}%
  \BibitemOpen
  \bibfield  {author} {\bibinfo {author} {\bibfnamefont {J.~L.}\ \bibnamefont
  {Vassent}}, \bibinfo {author} {\bibfnamefont {M.}~\bibnamefont {Dynna}},
  \bibinfo {author} {\bibfnamefont {A.}~\bibnamefont {Marty}}, \bibinfo
  {author} {\bibfnamefont {B.}~\bibnamefont {Gilles}}, \ and\ \bibinfo {author}
  {\bibfnamefont {G.}~\bibnamefont {Patrat}},\ }\Doi {10.1063/1.363626}
  {\bibfield  {journal} {\bibinfo  {journal} {Journal of Applied Physics},\
  }\textbf {\bibinfo {volume} {80}},\ \bibinfo {pages} {5727} (\bibinfo {year}
  {1996})}\BibitemShut {NoStop}%
\bibitem [{\citenamefont {Dynna}\ \emph {et~al.}(1996)\citenamefont {Dynna},
  \citenamefont {Vassent}, \citenamefont {Marty},\ and\ \citenamefont
  {Gilles}}]{Dynna96}%
  \BibitemOpen
  \bibfield  {author} {\bibinfo {author} {\bibfnamefont {M.}~\bibnamefont
  {Dynna}}, \bibinfo {author} {\bibfnamefont {J.~L.}\ \bibnamefont {Vassent}},
  \bibinfo {author} {\bibfnamefont {A.}~\bibnamefont {Marty}}, \ and\ \bibinfo
  {author} {\bibfnamefont {B.}~\bibnamefont {Gilles}},\ }\Doi
  {10.1063/1.363181} {\bibfield  {journal} {\bibinfo  {journal} {Journal of
  Applied Physics},\ }\textbf {\bibinfo {volume} {80}},\ \bibinfo {pages}
  {2650} (\bibinfo {year} {1996})}\BibitemShut {NoStop}%
\bibitem [{\citenamefont {Wulfhekel}\ \emph {et~al.}(2001)\citenamefont
  {Wulfhekel}, \citenamefont {Klaua}, \citenamefont {Ullmann}, \citenamefont
  {Zavaliche}, \citenamefont {Kirschner}, \citenamefont {Urban}, \citenamefont
  {Monchesky},\ and\ \citenamefont {Heinrich}}]{Wulfhekel01}%
  \BibitemOpen
  \bibfield  {author} {\bibinfo {author} {\bibfnamefont {W.}~\bibnamefont
  {Wulfhekel}}, \bibinfo {author} {\bibfnamefont {M.}~\bibnamefont {Klaua}},
  \bibinfo {author} {\bibfnamefont {D.}~\bibnamefont {Ullmann}}, \bibinfo
  {author} {\bibfnamefont {F.}~\bibnamefont {Zavaliche}}, \bibinfo {author}
  {\bibfnamefont {J.}~\bibnamefont {Kirschner}}, \bibinfo {author}
  {\bibfnamefont {R.}~\bibnamefont {Urban}}, \bibinfo {author} {\bibfnamefont
  {T.}~\bibnamefont {Monchesky}}, \ and\ \bibinfo {author} {\bibfnamefont
  {B.}~\bibnamefont {Heinrich}},\ }\Doi {10.1063/1.1342778} {\bibfield
  {journal} {\bibinfo  {journal} {Applied Physics Letters},\ }\textbf {\bibinfo
  {volume} {78}},\ \bibinfo {pages} {509} (\bibinfo {year} {2001})}\BibitemShut
  {NoStop}%
\bibitem [{\citenamefont {Wang}\ \emph {et~al.}(2005)\citenamefont {Wang},
  \citenamefont {Przybylski}, \citenamefont {Kuch}, \citenamefont {Chelaru},
  \citenamefont {Wang}, \citenamefont {Lu}, \citenamefont {Barthel},
  \citenamefont {Meyerheim},\ and\ \citenamefont {Kirschner}}]{Wang05a}%
  \BibitemOpen
  \bibfield  {author} {\bibinfo {author} {\bibfnamefont {W.}~\bibnamefont
  {Wang}}, \bibinfo {author} {\bibfnamefont {M.}~\bibnamefont {Przybylski}},
  \bibinfo {author} {\bibfnamefont {W.}~\bibnamefont {Kuch}}, \bibinfo {author}
  {\bibfnamefont {L.}~\bibnamefont {Chelaru}}, \bibinfo {author} {\bibfnamefont
  {J.}~\bibnamefont {Wang}}, \bibinfo {author} {\bibfnamefont {Y.}~\bibnamefont
  {Lu}}, \bibinfo {author} {\bibfnamefont {J.}~\bibnamefont {Barthel}},
  \bibinfo {author} {\bibfnamefont {H.}~\bibnamefont {Meyerheim}}, \ and\
  \bibinfo {author} {\bibfnamefont {J.}~\bibnamefont {Kirschner}},\ }\Doi
  {10.1103/PhysRevB.71.144416} {\bibfield  {journal} {\bibinfo  {journal}
  {Phys. Rev. B},\ }\textbf {\bibinfo {volume} {71}} (\bibinfo {year}
  {2005})},\ ISSN \bibinfo {issn} {1098-0121},\ \doi
  {10.1103/PhysRevB.71.144416}\BibitemShut {NoStop}%
\bibitem [{\citenamefont {W\"ustenberg}\ \emph {et~al.}(2012)\citenamefont
  {W\"ustenberg}, \citenamefont {Fetzer}, \citenamefont {Aeschlimann},
  \citenamefont {Cinchetti}, \citenamefont {Min\'ar}, \citenamefont {Braun},
  \citenamefont {Ebert}, \citenamefont {Ishikawa}, \citenamefont {Uemura},\
  and\ \citenamefont {Yamamoto}}]{Wuestenberg12}%
  \BibitemOpen
  \bibfield  {author} {\bibinfo {author} {\bibfnamefont {J.-P.}\ \bibnamefont
  {W\"ustenberg}}, \bibinfo {author} {\bibfnamefont {R.}~\bibnamefont
  {Fetzer}}, \bibinfo {author} {\bibfnamefont {M.}~\bibnamefont {Aeschlimann}},
  \bibinfo {author} {\bibfnamefont {M.}~\bibnamefont {Cinchetti}}, \bibinfo
  {author} {\bibfnamefont {J.}~\bibnamefont {Min\'ar}}, \bibinfo {author}
  {\bibfnamefont {J.}~\bibnamefont {Braun}}, \bibinfo {author} {\bibfnamefont
  {H.}~\bibnamefont {Ebert}}, \bibinfo {author} {\bibfnamefont
  {T.}~\bibnamefont {Ishikawa}}, \bibinfo {author} {\bibfnamefont
  {T.}~\bibnamefont {Uemura}}, \ and\ \bibinfo {author} {\bibfnamefont
  {M.}~\bibnamefont {Yamamoto}},\ }\Doi {10.1103/PhysRevB.85.064407} {\bibfield
   {journal} {\bibinfo  {journal} {Phys. Rev. B},\ }\textbf {\bibinfo {volume}
  {85}},\ \bibinfo {pages} {064407} (\bibinfo {year} {2012})}\BibitemShut
  {NoStop}%
\bibitem [{\citenamefont {Cinchetti}\ \emph {et~al.}(2007)\citenamefont
  {Cinchetti}, \citenamefont {W\"ustenberg}, \citenamefont {Albaneda},
  \citenamefont {Steeb}, \citenamefont {Conca}, \citenamefont {Jourdan},\ and\
  \citenamefont {Aeschlimann}}]{Cinchetti07}%
  \BibitemOpen
  \bibfield  {author} {\bibinfo {author} {\bibfnamefont {M.}~\bibnamefont
  {Cinchetti}}, \bibinfo {author} {\bibfnamefont {J.-P.}\ \bibnamefont
  {W\"ustenberg}}, \bibinfo {author} {\bibfnamefont {M.~S.}\ \bibnamefont
  {Albaneda}}, \bibinfo {author} {\bibfnamefont {F.}~\bibnamefont {Steeb}},
  \bibinfo {author} {\bibfnamefont {A.}~\bibnamefont {Conca}}, \bibinfo
  {author} {\bibfnamefont {M.}~\bibnamefont {Jourdan}}, \ and\ \bibinfo
  {author} {\bibfnamefont {M.}~\bibnamefont {Aeschlimann}},\ }\Doi
  {10.1088/0022-3727/40/6/S05} {\bibfield  {journal} {\bibinfo  {journal} {J.
  Phys. D: Appl. Phys.},\ }\textbf {\bibinfo {volume} {40}},\ \bibinfo {pages}
  {1544} (\bibinfo {year} {2007})}\BibitemShut {NoStop}%
\bibitem [{\citenamefont {Schneider}\ \emph {et~al.}(2006)\citenamefont
  {Schneider}, \citenamefont {Jakob}, \citenamefont {Kallmayer}, \citenamefont
  {Elmers}, \citenamefont {Cinchetti}, \citenamefont {Balke}, \citenamefont
  {Wurmehl}, \citenamefont {Felser}, \citenamefont {Aeschlimann},\ and\
  \citenamefont {Adrian}}]{Schneider06}%
  \BibitemOpen
  \bibfield  {author} {\bibinfo {author} {\bibfnamefont {H.}~\bibnamefont
  {Schneider}}, \bibinfo {author} {\bibfnamefont {G.}~\bibnamefont {Jakob}},
  \bibinfo {author} {\bibfnamefont {M.}~\bibnamefont {Kallmayer}}, \bibinfo
  {author} {\bibfnamefont {H.~J.}\ \bibnamefont {Elmers}}, \bibinfo {author}
  {\bibfnamefont {M.}~\bibnamefont {Cinchetti}}, \bibinfo {author}
  {\bibfnamefont {B.}~\bibnamefont {Balke}}, \bibinfo {author} {\bibfnamefont
  {S.}~\bibnamefont {Wurmehl}}, \bibinfo {author} {\bibfnamefont
  {C.}~\bibnamefont {Felser}}, \bibinfo {author} {\bibfnamefont
  {M.}~\bibnamefont {Aeschlimann}}, \ and\ \bibinfo {author} {\bibfnamefont
  {H.}~\bibnamefont {Adrian}},\ }\Doi {10.1103/PhysRevB.74.174426} {\bibfield
  {journal} {\bibinfo  {journal} {Phys. Rev. B},\ }\textbf {\bibinfo {volume}
  {74}},\ \bibinfo {eid} {174426} (\bibinfo {year} {2006})}\BibitemShut
  {NoStop}%
\bibitem [{\citenamefont {W\"ustenberg}\ \emph {et~al.}(2009)\citenamefont
  {W\"ustenberg}, \citenamefont {Fischer}, \citenamefont {Herbort},
  \citenamefont {Jourdan}, \citenamefont {Aeschlimann},\ and\ \citenamefont
  {Cinchetti}}]{Wuestenberg09}%
  \BibitemOpen
  \bibfield  {author} {\bibinfo {author} {\bibfnamefont {J.-P.}\ \bibnamefont
  {W\"ustenberg}}, \bibinfo {author} {\bibfnamefont {J.}~\bibnamefont
  {Fischer}}, \bibinfo {author} {\bibfnamefont {C.}~\bibnamefont {Herbort}},
  \bibinfo {author} {\bibfnamefont {M.}~\bibnamefont {Jourdan}}, \bibinfo
  {author} {\bibfnamefont {M.}~\bibnamefont {Aeschlimann}}, \ and\ \bibinfo
  {author} {\bibfnamefont {M.}~\bibnamefont {Cinchetti}},\ }\Doi
  {10.1088/0022-3727/42/8/084016} {\bibfield  {journal} {\bibinfo  {journal}
  {J. Phys. D: Appl. Phys.},\ }\textbf {\bibinfo {volume} {42}},\ \bibinfo
  {pages} {084016} (\bibinfo {year} {2009})}\BibitemShut {NoStop}%
\bibitem [{\citenamefont {Kolbe}\ \emph {et~al.}(2012)\citenamefont {Kolbe},
  \citenamefont {Chadov}, \citenamefont {Jorge}, \citenamefont {Sch\"onhense},
  \citenamefont {Felser}, \citenamefont {Elmers}, \citenamefont {Kl\"aui},\
  and\ \citenamefont {Jourdan}}]{Kolbe12}%
  \BibitemOpen
  \bibfield  {author} {\bibinfo {author} {\bibfnamefont {M.}~\bibnamefont
  {Kolbe}}, \bibinfo {author} {\bibfnamefont {S.}~\bibnamefont {Chadov}},
  \bibinfo {author} {\bibfnamefont {E.~A.}\ \bibnamefont {Jorge}}, \bibinfo
  {author} {\bibfnamefont {G.}~\bibnamefont {Sch\"onhense}}, \bibinfo {author}
  {\bibfnamefont {C.}~\bibnamefont {Felser}}, \bibinfo {author} {\bibfnamefont
  {H.-J.}\ \bibnamefont {Elmers}}, \bibinfo {author} {\bibfnamefont
  {M.}~\bibnamefont {Kl\"aui}}, \ and\ \bibinfo {author} {\bibfnamefont
  {M.}~\bibnamefont {Jourdan}},\ }\Doi {10.1103/PhysRevB.86.024422} {\bibfield
  {journal} {\bibinfo  {journal} {Phys. Rev. B},\ }\textbf {\bibinfo {volume}
  {86}},\ \bibinfo {pages} {024422} (\bibinfo {year} {2012})}\BibitemShut
  {NoStop}%
\bibitem [{\citenamefont {Sicot}\ \emph {et~al.}(2003)\citenamefont {Sicot},
  \citenamefont {Andrieu}, \citenamefont {Turban}, \citenamefont
  {Fagot-Revurat}, \citenamefont {Cercellier}, \citenamefont {Tagliaferri},
  \citenamefont {De~Nadai}, \citenamefont {Brookes}, \citenamefont {Bertran},\
  and\ \citenamefont {Fortuna}}]{Sicot03}%
  \BibitemOpen
  \bibfield  {author} {\bibinfo {author} {\bibfnamefont {M.}~\bibnamefont
  {Sicot}}, \bibinfo {author} {\bibfnamefont {S.}~\bibnamefont {Andrieu}},
  \bibinfo {author} {\bibfnamefont {P.}~\bibnamefont {Turban}}, \bibinfo
  {author} {\bibfnamefont {Y.}~\bibnamefont {Fagot-Revurat}}, \bibinfo {author}
  {\bibfnamefont {H.}~\bibnamefont {Cercellier}}, \bibinfo {author}
  {\bibfnamefont {A.}~\bibnamefont {Tagliaferri}}, \bibinfo {author}
  {\bibfnamefont {C.}~\bibnamefont {De~Nadai}}, \bibinfo {author}
  {\bibfnamefont {N.~B.}\ \bibnamefont {Brookes}}, \bibinfo {author}
  {\bibfnamefont {F.}~\bibnamefont {Bertran}}, \ and\ \bibinfo {author}
  {\bibfnamefont {F.}~\bibnamefont {Fortuna}},\ }\Doi
  {10.1103/PhysRevB.68.184406} {\bibfield  {journal} {\bibinfo  {journal}
  {Phys. Rev. B},\ }\textbf {\bibinfo {volume} {68}},\ \bibinfo {pages}
  {184406} (\bibinfo {year} {2003})}\BibitemShut {NoStop}%
\bibitem [{\citenamefont {Matthes}\ \emph {et~al.}(2004)\citenamefont
  {Matthes}, \citenamefont {Tong},\ and\ \citenamefont
  {Schneider}}]{Matthes04}%
  \BibitemOpen
  \bibfield  {author} {\bibinfo {author} {\bibfnamefont {F.}~\bibnamefont
  {Matthes}}, \bibinfo {author} {\bibfnamefont {L.-N.}\ \bibnamefont {Tong}}, \
  and\ \bibinfo {author} {\bibfnamefont {C.~M.}\ \bibnamefont {Schneider}},\
  }\Doi {10.1063/1.1669214} {\bibfield  {journal} {\bibinfo  {journal} {Journal
  of Applied Physics},\ }\textbf {\bibinfo {volume} {95}},\ \bibinfo {pages}
  {7240} (\bibinfo {year} {2004})}\BibitemShut {NoStop}%
\bibitem [{\citenamefont {M\"uller}\ \emph {et~al.}(2007)\citenamefont
  {M\"uller}, \citenamefont {Matthes},\ and\ \citenamefont
  {Schneider}}]{Mueller07}%
  \BibitemOpen
  \bibfield  {author} {\bibinfo {author} {\bibfnamefont {M.}~\bibnamefont
  {M\"uller}}, \bibinfo {author} {\bibfnamefont {F.}~\bibnamefont {Matthes}}, \
  and\ \bibinfo {author} {\bibfnamefont {C.~M.}\ \bibnamefont {Schneider}},\
  }\Doi {10.1063/1.2711418} {\bibfield  {journal} {\bibinfo  {journal} {Journal
  of Applied Physics},\ }\textbf {\bibinfo {volume} {101}},\ \bibinfo {eid}
  {09G519} (\bibinfo {year} {2007})}\BibitemShut {NoStop}%
\bibitem [{\citenamefont {Tong}\ \emph {et~al.}(2006)\citenamefont {Tong},
  \citenamefont {Deng}, \citenamefont {Matthes}, \citenamefont {M\"uller},
  \citenamefont {Schneider},\ and\ \citenamefont {Lee}}]{Tong06}%
  \BibitemOpen
  \bibfield  {author} {\bibinfo {author} {\bibfnamefont {L.-N.}\ \bibnamefont
  {Tong}}, \bibinfo {author} {\bibfnamefont {C.-L.}\ \bibnamefont {Deng}},
  \bibinfo {author} {\bibfnamefont {F.}~\bibnamefont {Matthes}}, \bibinfo
  {author} {\bibfnamefont {M.}~\bibnamefont {M\"uller}}, \bibinfo {author}
  {\bibfnamefont {C.~M.}\ \bibnamefont {Schneider}}, \ and\ \bibinfo {author}
  {\bibfnamefont {C.-G.}\ \bibnamefont {Lee}},\ }\Doi
  {10.1103/PhysRevB.73.214401} {\bibfield  {journal} {\bibinfo  {journal}
  {Phys. Rev. B},\ }\textbf {\bibinfo {volume} {73}},\ \bibinfo {pages}
  {214401} (\bibinfo {year} {2006})}\BibitemShut {NoStop}%
\bibitem [{\citenamefont {Bonell}\ \emph {et~al.}(2012)\citenamefont {Bonell},
  \citenamefont {Hauet}, \citenamefont {Andrieu}, \citenamefont {Bertran},
  \citenamefont {Le~F\`evre}, \citenamefont {Calmels}, \citenamefont {Tejeda},
  \citenamefont {Montaigne}, \citenamefont {Warot-Fonrose}, \citenamefont
  {Belhadji}, \citenamefont {Nicolaou},\ and\ \citenamefont
  {Taleb-Ibrahimi}}]{Bonell12}%
  \BibitemOpen
  \bibfield  {author} {\bibinfo {author} {\bibfnamefont {F.}~\bibnamefont
  {Bonell}}, \bibinfo {author} {\bibfnamefont {T.}~\bibnamefont {Hauet}},
  \bibinfo {author} {\bibfnamefont {S.}~\bibnamefont {Andrieu}}, \bibinfo
  {author} {\bibfnamefont {F.}~\bibnamefont {Bertran}}, \bibinfo {author}
  {\bibfnamefont {P.}~\bibnamefont {Le~F\`evre}}, \bibinfo {author}
  {\bibfnamefont {L.}~\bibnamefont {Calmels}}, \bibinfo {author} {\bibfnamefont
  {A.}~\bibnamefont {Tejeda}}, \bibinfo {author} {\bibfnamefont
  {F.}~\bibnamefont {Montaigne}}, \bibinfo {author} {\bibfnamefont
  {B.}~\bibnamefont {Warot-Fonrose}}, \bibinfo {author} {\bibfnamefont
  {B.}~\bibnamefont {Belhadji}}, \bibinfo {author} {\bibfnamefont
  {A.}~\bibnamefont {Nicolaou}}, \ and\ \bibinfo {author} {\bibfnamefont
  {A.}~\bibnamefont {Taleb-Ibrahimi}},\ }\Doi {10.1103/PhysRevLett.108.176602}
  {\bibfield  {journal} {\bibinfo  {journal} {Phys. Rev. Lett.},\ }\textbf
  {\bibinfo {volume} {108}},\ \bibinfo {pages} {176602} (\bibinfo {year}
  {2012})}\BibitemShut {NoStop}%
\bibitem [{\citenamefont {Plucinski}\ \emph {et~al.}(2007)\citenamefont
  {Plucinski}, \citenamefont {Zhao}, \citenamefont {Sinkovic},\ and\
  \citenamefont {Vescovo}}]{Plucinski07}%
  \BibitemOpen
  \bibfield  {author} {\bibinfo {author} {\bibfnamefont {L.}~\bibnamefont
  {Plucinski}}, \bibinfo {author} {\bibfnamefont {Y.}~\bibnamefont {Zhao}},
  \bibinfo {author} {\bibfnamefont {B.}~\bibnamefont {Sinkovic}}, \ and\
  \bibinfo {author} {\bibfnamefont {E.}~\bibnamefont {Vescovo}},\ }\Doi
  {10.1103/PhysRevB.75.214411} {\bibfield  {journal} {\bibinfo  {journal}
  {Phys. Rev. B},\ }\textbf {\bibinfo {volume} {75}},\ \bibinfo {pages}
  {214411} (\bibinfo {year} {2007})}\BibitemShut {NoStop}%
\bibitem [{\citenamefont {Fetzer}\ \emph {et~al.}(2012)\citenamefont {Fetzer},
  \citenamefont {L\"osch}, \citenamefont {Ohdaira}, \citenamefont {Naganuma},
  \citenamefont {Oogane}, \citenamefont {Ando}, \citenamefont {Taira},
  \citenamefont {Uemura}, \citenamefont {Yamamoto}, \citenamefont
  {Aeschlimann},\ and\ \citenamefont {Cinchetti}}]{Fetzer12b}%
  \BibitemOpen
  \bibfield  {author} {\bibinfo {author} {\bibfnamefont {R.}~\bibnamefont
  {Fetzer}}, \bibinfo {author} {\bibfnamefont {M.}~\bibnamefont {L\"osch}},
  \bibinfo {author} {\bibfnamefont {Y.}~\bibnamefont {Ohdaira}}, \bibinfo
  {author} {\bibfnamefont {H.}~\bibnamefont {Naganuma}}, \bibinfo {author}
  {\bibfnamefont {M.}~\bibnamefont {Oogane}}, \bibinfo {author} {\bibfnamefont
  {Y.}~\bibnamefont {Ando}}, \bibinfo {author} {\bibfnamefont {T.}~\bibnamefont
  {Taira}}, \bibinfo {author} {\bibfnamefont {T.}~\bibnamefont {Uemura}},
  \bibinfo {author} {\bibfnamefont {M.}~\bibnamefont {Yamamoto}}, \bibinfo
  {author} {\bibfnamefont {M.}~\bibnamefont {Aeschlimann}}, \ and\ \bibinfo
  {author} {\bibfnamefont {M.}~\bibnamefont {Cinchetti}},\ }\href@noop {}
  {\bibfield  {journal} {\bibinfo  {journal} {arXiv:1209.4368}} (\bibinfo
  {year} {2012})}\BibitemShut {NoStop}%
\bibitem [{\citenamefont {Yamamoto}\ \emph {et~al.}(2010)\citenamefont
  {Yamamoto}, \citenamefont {Ishikawa}, \citenamefont {Taira}, \citenamefont
  {fang Li}, \citenamefont {ichi Matsuda},\ and\ \citenamefont
  {Uemura}}]{Yamamoto10}%
  \BibitemOpen
  \bibfield  {author} {\bibinfo {author} {\bibfnamefont {M.}~\bibnamefont
  {Yamamoto}}, \bibinfo {author} {\bibfnamefont {T.}~\bibnamefont {Ishikawa}},
  \bibinfo {author} {\bibfnamefont {T.}~\bibnamefont {Taira}}, \bibinfo
  {author} {\bibfnamefont {G.}~\bibnamefont {fang Li}}, \bibinfo {author}
  {\bibfnamefont {K.}~\bibnamefont {ichi Matsuda}}, \ and\ \bibinfo {author}
  {\bibfnamefont {T.}~\bibnamefont {Uemura}},\ }\href
  {http://stacks.iop.org/0953-8984/22/i=16/a=164212} {\bibfield  {journal}
  {\bibinfo  {journal} {Journal of Physics: Condensed Matter},\ }\textbf
  {\bibinfo {volume} {22}},\ \bibinfo {pages} {164212} (\bibinfo {year}
  {2010})}\BibitemShut {NoStop}%
\bibitem [{\citenamefont {Davis}\ \emph {et~al.}(1987)\citenamefont {Davis},
  \citenamefont {MacDonald}, \citenamefont {Palmberg},\ and\ \citenamefont
  {Riach}}]{Davis87}%
  \BibitemOpen
  \bibfield  {author} {\bibinfo {author} {\bibfnamefont {L.}~\bibnamefont
  {Davis}}, \bibinfo {author} {\bibfnamefont {N.}~\bibnamefont {MacDonald}},
  \bibinfo {author} {\bibfnamefont {P.}~\bibnamefont {Palmberg}}, \ and\
  \bibinfo {author} {\bibfnamefont {G.}~\bibnamefont {Riach}},\ }\href@noop {}
  {\emph {\bibinfo {title} {Handbook of Auger Electron Spectroscopy: A
  Reference Book of Standard Data for Identification and Interpretation of
  Auger Electron Spectroscopy Data}}}\ (\bibinfo  {publisher} {Physical
  Electronics},\ \bibinfo {year} {1987})\BibitemShut {NoStop}%
\bibitem [{\citenamefont {C.J.~Powell}(2000)}]{Nist00}%
  \BibitemOpen
  \bibfield  {author} {\bibinfo {author} {\bibfnamefont {A.}~\bibnamefont
  {C.J.~Powell}},\ }\href@noop {} {\emph {\bibinfo {title} {{NIST Electron
  Inelastic-Mean-Free-Path Database}}}},\ \bibinfo {type} {Tech. Rep.}\
  (\bibinfo  {institution} {National Institute of Standards and Technology},\
  \bibinfo {year} {2000})\BibitemShut {NoStop}%
\bibitem [{\citenamefont {Henrich}(1985)}]{Henrich85}%
  \BibitemOpen
  \bibfield  {author} {\bibinfo {author} {\bibfnamefont {V.}~\bibnamefont
  {Henrich}},\ }\href@noop {} {\bibfield  {journal} {\bibinfo  {journal}
  {Reports on Progress in Physics},\ }\textbf {\bibinfo {volume} {48}},\
  \bibinfo {pages} {1481} (\bibinfo {year} {1985})}\BibitemShut {NoStop}%
\bibitem [{\citenamefont {Geneste}\ \emph {et~al.}(2005)\citenamefont
  {Geneste}, \citenamefont {Morillo},\ and\ \citenamefont
  {Finocchi}}]{Geneste05}%
  \BibitemOpen
  \bibfield  {author} {\bibinfo {author} {\bibfnamefont {G.}~\bibnamefont
  {Geneste}}, \bibinfo {author} {\bibfnamefont {J.}~\bibnamefont {Morillo}}, \
  and\ \bibinfo {author} {\bibfnamefont {F.}~\bibnamefont {Finocchi}},\ }\Doi
  {10.1063/1.1886734} {\bibfield  {journal} {\bibinfo  {journal} {The Journal
  of Chemical Physics},\ }\textbf {\bibinfo {volume} {122}},\ \bibinfo {eid}
  {174707} (\bibinfo {year} {2005})}\BibitemShut {NoStop}%
\bibitem [{\citenamefont {Ochs}\ \emph {et~al.}(1998)\citenamefont {Ochs},
  \citenamefont {Brause}, \citenamefont {Braun}, \citenamefont
  {Maus-Friedrichs},\ and\ \citenamefont {Kempter}}]{Ochs98}%
  \BibitemOpen
  \bibfield  {author} {\bibinfo {author} {\bibfnamefont {D.}~\bibnamefont
  {Ochs}}, \bibinfo {author} {\bibfnamefont {M.}~\bibnamefont {Brause}},
  \bibinfo {author} {\bibfnamefont {B.}~\bibnamefont {Braun}}, \bibinfo
  {author} {\bibfnamefont {W.}~\bibnamefont {Maus-Friedrichs}}, \ and\ \bibinfo
  {author} {\bibfnamefont {V.}~\bibnamefont {Kempter}},\ }\href
  {http://www.sciencedirect.com/science/article/pii/S003960289700722X}
  {\bibfield  {journal} {\bibinfo  {journal} {Surface Science},\ }\textbf
  {\bibinfo {volume} {397}},\ \bibinfo {pages} {101} (\bibinfo {year}
  {1998})},\ ISSN \bibinfo {issn} {0039-6028}\BibitemShut {NoStop}%
\bibitem [{\citenamefont {Pacchioni}\ \emph {et~al.}(1994)\citenamefont
  {Pacchioni}, \citenamefont {Ricart},\ and\ \citenamefont
  {Illas}}]{Pacchioni94}%
  \BibitemOpen
  \bibfield  {author} {\bibinfo {author} {\bibfnamefont {G.}~\bibnamefont
  {Pacchioni}}, \bibinfo {author} {\bibfnamefont {J.~M.}\ \bibnamefont
  {Ricart}}, \ and\ \bibinfo {author} {\bibfnamefont {F.}~\bibnamefont
  {Illas}},\ }\Doi {10.1021/ja00101a038} {\bibfield  {journal} {\bibinfo
  {journal} {Journal of the American Chemical Society},\ }\textbf {\bibinfo
  {volume} {116}},\ \bibinfo {pages} {10152} (\bibinfo {year} {1994})},\
  \Eprint
  {http://arxiv.org/abs/http://pubs.acs.org/doi/pdf/10.1021/ja00101a038}
  {http://pubs.acs.org/doi/pdf/10.1021/ja00101a038} \BibitemShut {NoStop}%
\bibitem [{\citenamefont {Pacchioni}(1993)}]{Pacchioni93}%
  \BibitemOpen
  \bibfield  {author} {\bibinfo {author} {\bibfnamefont {G.}~\bibnamefont
  {Pacchioni}},\ }\href
  {http://www.sciencedirect.com/science/article/pii/003960289390869L}
  {\bibfield  {journal} {\bibinfo  {journal} {Surface Science},\ }\textbf
  {\bibinfo {volume} {281}},\ \bibinfo {pages} {207} (\bibinfo {year}
  {1993})},\ ISSN \bibinfo {issn} {0039-6028}\BibitemShut {NoStop}%
\bibitem [{\citenamefont {Pacchioni}\ \emph {et~al.}(1992)\citenamefont
  {Pacchioni}, \citenamefont {Minerva},\ and\ \citenamefont
  {Bagus}}]{Pacchioni92}%
  \BibitemOpen
  \bibfield  {author} {\bibinfo {author} {\bibfnamefont {G.}~\bibnamefont
  {Pacchioni}}, \bibinfo {author} {\bibfnamefont {T.}~\bibnamefont {Minerva}},
  \ and\ \bibinfo {author} {\bibfnamefont {P.~S.}\ \bibnamefont {Bagus}},\
  }\href {http://www.sciencedirect.com/science/article/pii/003960289290818Q}
  {\bibfield  {journal} {\bibinfo  {journal} {Surface Science},\ }\textbf
  {\bibinfo {volume} {275}},\ \bibinfo {pages} {450} (\bibinfo {year}
  {1992})},\ ISSN \bibinfo {issn} {0039-6028}\BibitemShut {NoStop}%
\bibitem [{\citenamefont {L.~N.~Kantorovich}\ and\ \citenamefont
  {White}(1996)}]{Kantorovich96}%
  \BibitemOpen
  \bibfield  {author} {\bibinfo {author} {\bibfnamefont {M.~J.~G.}\
  \bibnamefont {L.~N.~Kantorovich}}\ and\ \bibinfo {author} {\bibfnamefont
  {J.~A.}\ \bibnamefont {White}},\ }\href@noop {} {\bibfield  {journal}
  {\bibinfo  {journal} {Journal of the Chemical Society, Faraday
  Transactions},\ }\textbf {\bibinfo {volume} {92}},\ \bibinfo {pages} {2075}
  (\bibinfo {year} {1996})}\BibitemShut {NoStop}%
\bibitem [{\citenamefont {Kantorovich}\ and\ \citenamefont
  {Gillan}(1997)}]{Kantorovich97}%
  \BibitemOpen
  \bibfield  {author} {\bibinfo {author} {\bibfnamefont {L.}~\bibnamefont
  {Kantorovich}}\ and\ \bibinfo {author} {\bibfnamefont {M.}~\bibnamefont
  {Gillan}},\ }\href
  {http://www.sciencedirect.com/science/article/pii/S0039602896012447}
  {\bibfield  {journal} {\bibinfo  {journal} {Surface Science},\ }\textbf
  {\bibinfo {volume} {374}},\ \bibinfo {pages} {373} (\bibinfo {year}
  {1997})},\ ISSN \bibinfo {issn} {0039-6028}\BibitemShut {NoStop}%
\bibitem [{\citenamefont {Wu}\ \emph {et~al.}(1991)\citenamefont {Wu},
  \citenamefont {Corneille}, \citenamefont {Estrada}, \citenamefont {He},\ and\
  \citenamefont {Wayne~Goodman}}]{Wu91}%
  \BibitemOpen
  \bibfield  {author} {\bibinfo {author} {\bibfnamefont {M.-C.}\ \bibnamefont
  {Wu}}, \bibinfo {author} {\bibfnamefont {J.~S.}\ \bibnamefont {Corneille}},
  \bibinfo {author} {\bibfnamefont {C.~A.}\ \bibnamefont {Estrada}}, \bibinfo
  {author} {\bibfnamefont {J.-W.}\ \bibnamefont {He}}, \ and\ \bibinfo {author}
  {\bibfnamefont {D.}~\bibnamefont {Wayne~Goodman}},\ }\href
  {http://www.sciencedirect.com/science/article/pii/000926149190110U}
  {\bibfield  {journal} {\bibinfo  {journal} {Chemical Physics Letters},\
  }\textbf {\bibinfo {volume} {182}},\ \bibinfo {pages} {472} (\bibinfo {year}
  {1991})},\ ISSN \bibinfo {issn} {0009-2614}\BibitemShut {NoStop}%
\bibitem [{\citenamefont {Tegenkamp}\ \emph {et~al.}(1999)\citenamefont
  {Tegenkamp}, \citenamefont {Pfn\"ur}, \citenamefont {Ernst}, \citenamefont
  {Malaske}, \citenamefont {Wollschl\"ager}, \citenamefont {Peterka},
  \citenamefont {Schr\"oder}, \citenamefont {Zielasek},\ and\ \citenamefont
  {Henzler}}]{Tegenkamp99}%
  \BibitemOpen
  \bibfield  {author} {\bibinfo {author} {\bibfnamefont {C.}~\bibnamefont
  {Tegenkamp}}, \bibinfo {author} {\bibfnamefont {H.}~\bibnamefont {Pfn\"ur}},
  \bibinfo {author} {\bibfnamefont {W.}~\bibnamefont {Ernst}}, \bibinfo
  {author} {\bibfnamefont {U.}~\bibnamefont {Malaske}}, \bibinfo {author}
  {\bibfnamefont {J.}~\bibnamefont {Wollschl\"ager}}, \bibinfo {author}
  {\bibfnamefont {D.}~\bibnamefont {Peterka}}, \bibinfo {author} {\bibfnamefont
  {K.~M.}\ \bibnamefont {Schr\"oder}}, \bibinfo {author} {\bibfnamefont
  {V.}~\bibnamefont {Zielasek}}, \ and\ \bibinfo {author} {\bibfnamefont
  {M.}~\bibnamefont {Henzler}},\ }\href
  {http://stacks.iop.org/0953-8984/11/i=49/a=312} {\bibfield  {journal}
  {\bibinfo  {journal} {Journal of Physics: Condensed Matter},\ }\textbf
  {\bibinfo {volume} {11}},\ \bibinfo {pages} {9943} (\bibinfo {year}
  {1999})}\BibitemShut {NoStop}%
\bibitem [{\citenamefont {Miura}\ \emph {et~al.}(2008)\citenamefont {Miura},
  \citenamefont {Uchida}, \citenamefont {Oba}, \citenamefont {Abe},\ and\
  \citenamefont {Shirai}}]{Miura08}%
  \BibitemOpen
  \bibfield  {author} {\bibinfo {author} {\bibfnamefont {Y.}~\bibnamefont
  {Miura}}, \bibinfo {author} {\bibfnamefont {H.}~\bibnamefont {Uchida}},
  \bibinfo {author} {\bibfnamefont {Y.}~\bibnamefont {Oba}}, \bibinfo {author}
  {\bibfnamefont {K.}~\bibnamefont {Abe}}, \ and\ \bibinfo {author}
  {\bibfnamefont {M.}~\bibnamefont {Shirai}},\ }\Doi
  {10.1103/PhysRevB.78.064416} {\bibfield  {journal} {\bibinfo  {journal}
  {Phys. Rev. B},\ }\textbf {\bibinfo {volume} {78}},\ \bibinfo {pages}
  {064416} (\bibinfo {year} {2008})}\BibitemShut {NoStop}%
\bibitem [{\citenamefont {Hermanson}(1977)}]{Hermanson77}%
  \BibitemOpen
  \bibfield  {author} {\bibinfo {author} {\bibfnamefont {J.}~\bibnamefont
  {Hermanson}},\ }\Doi {10.1016/0038-1098(77)90931-0} {\bibfield  {journal}
  {\bibinfo  {journal} {Solid State Communications},\ }\textbf {\bibinfo
  {volume} {22}},\ \bibinfo {pages} {9 } (\bibinfo {year} {1977})},\ ISSN
  \bibinfo {issn} {0038-1098}\BibitemShut {NoStop}%
\bibitem [{\citenamefont {Eberhardt}\ and\ \citenamefont
  {Himpsel}(1980)}]{Eberhardt80}%
  \BibitemOpen
  \bibfield  {author} {\bibinfo {author} {\bibfnamefont {W.}~\bibnamefont
  {Eberhardt}}\ and\ \bibinfo {author} {\bibfnamefont {F.~J.}\ \bibnamefont
  {Himpsel}},\ }\Doi {10.1103/PhysRevB.21.5572} {\bibfield  {journal} {\bibinfo
   {journal} {Phys. Rev. B},\ }\textbf {\bibinfo {volume} {21}},\ \bibinfo
  {pages} {5572} (\bibinfo {year} {1980})}\BibitemShut {NoStop}%
\end{thebibliography}%

\end{document}